\documentclass[acmlarge, authorversion,nonacm]{acmart}

\AtBeginDocument{%
  }

\setcopyright{acmlicensed}
\copyrightyear{2024}
\acmYear{2024}
\acmDOI{XXXXXXX.XXXXXXX}
\usepackage{multirow}
\usepackage[capitalise]{cleveref}
\usepackage{subfigure,float}
\usepackage{algorithm}
\usepackage[noend]{algorithmic}

\acmJournal{IMWUT}
\acmVolume{0}
\acmNumber{0}
\acmArticle{0}
\acmMonth{5}

\acmSubmissionID{8900}

\begin{document}

\title{Unsupervised Statistical Feature-Guided Diffusion Model for Sensor-based Human Activity Recognition}

\author{Si Zuo}
\authornote{Both authors contributed equally to this research.}
\affiliation{%
  \institution{Aalto University}
  \city{Espoo}
  \country{Finland}
}

\author{Vitor Fortes Rey}
\authornotemark[1]
\affiliation{%
  \institution{DFKI and RPTU Kaiserslautern-Landau}
  \city{Kaiserslautern}
  \country{Germany}
}

\author{Sungho Suh}
\authornote{Corresponding author}
\affiliation{%
  \institution{DFKI and RPTU Kaiserslautern-Landau}
  \city{Kaiserslautern}
  \country{Germany}
}
\email{sungho.suh@dfki.de}

\author{Stephan Sigg}
\affiliation{%
  \institution{Aalto University}
  \city{Espoo}
  \country{Finland}
}

\author{Paul Lukowicz}
\affiliation{%
  \institution{DFKI and RPTU Kaiserslautern-Landau}
  \city{Kaiserslautern}
  \country{Germany}
}

\renewcommand{\shortauthors}{Zuo et al.}

\begin{abstract}
Human activity recognition (HAR) from on-body sensors is a core functionality in many AI applications: from personal health, through sports and wellness to Industry 4.0. A key problem holding up progress in wearable sensor-based HAR, compared to other ML areas, such as computer vision, is the unavailability of diverse and labeled training data. Particularly, while there are innumerable annotated images available in online repositories, freely available sensor data is sparse and mostly unlabeled. We propose an unsupervised statistical feature-guided diffusion model specifically optimized for wearable sensor-based human activity recognition with devices such as inertial measurement unit (IMU) sensors. The method generates synthetic labeled time-series sensor data without relying on annotated training data. Thereby, it addresses the scarcity and annotation difficulties associated with real-world sensor data. By conditioning the diffusion model on statistical information such as mean, standard deviation, Z-score, and skewness, we generate diverse and representative synthetic sensor data. We conducted experiments on public human activity recognition datasets and compared the method to conventional oversampling and state-of-the-art generative adversarial network methods. Experimental results demonstrate that this can improve the performance of human activity recognition and outperform existing techniques.
 \end{abstract}


\received{20 February 2007}
\received[revised]{12 March 2009}
\received[accepted]{5 June 2009}

\maketitle

\section{Introduction}

Wearable sensor-based human activity recognition (HAR) plays a crucial role in various domains, including healthcare~\cite{bachlin2009wearable,fischer2021masquare}, sports~\cite{sundholm2014smart,zhou2022quali}, and security~\cite{derawi2010unobtrusive}. 
The accurate recognition of human activities enables the development of intelligent systems and applications that can assist individuals, monitor their well-being, and improve safety. 
It is a core component of the human-centric AI vision.

While other areas of Machine Learning, such as Computer Vision or Natural Language Processing (NLP), have made dramatic progress over the last decade, facilitating a range of real-world applications, the performance of HAR systems is lagging behind. 
To a large extent, this is due to the shortage of labeled training data. 
Training data for HAR from wearable sensors is available in much smaller amounts than images, videos, or texts, and it is also more costly and challenging to annotate~\cite{fortes2022learning}. 
Unlike vision data, which can be annotated with tools that build on available image or video datasets~\cite{lin2019block,russell2008labelme}, wearable sensor data annotation is prohibitively expensive and time-consuming, which hinders the progress of wearable sensor-based HAR research and limits its practical applications. Moreover, time-series sensor data for HAR often lacks the rich semantic information that is abundant in computer vision data. At the same time, wearable sensor data offers advantages over visual information in many situations.
Unlike visual data, which may capture identifiable features or personal information, wearable sensor-based HAR data typically deals with raw measurements or abstract data that are not easily linked to an individual's identity. 
Compared to video or image, lighting conditions, and camera placement do not affect the quality of wearable sensor data. 
Other benefits include a smaller data size and less energy consumption. 
For these reasons, the popularity of Inertial Measurement Unit (IMU) sensors has been steadily increasing and it is found in mobile devices (e.g. smartwatches and smartphones), drones, robotics, motion capture, etc. 

Synthetic training data generation (and data set augmentation) is a solution to the lack of labeled data. 
Compared to Computer Vision and NLP, less work is focused on wearable sensor data generation. 
Furthermore, methods applied to Computer Vision and NLP are often not directly applicable to wearable sensor data.  
Consequently, in order to overcome the scarcity of annotated datasets for wearable sensor data, new effective methods are needed to generate labeled data.

To address this challenge, translating video data into IMU representations \cite{rey2019let,kwon2020imutube} has been studied. These efforts have leveraged generative methods \cite{rey2019let, fortes2021translating} and trajectory-based approaches \cite{kwon2020imutube, xiao2021deep} to extract IMU data from videos, expanding the applicability of sensor-based HAR. Despite the successes of HAR performance improvement, limitations exist, particularly regarding input video quality and vision-based limitations, such as camera ego-motion and object occlusion.

Conversely, traditional data augmentation techniques, common in various domains, have been adapted for wearable sensor data \cite{kim2021label, jeong2021sensor}. While computer vision relies on simple affine transformations, accelerometer signals require alternative strategies, such as signal processing methods like jittering and scaling \cite{um2017data, ohashi2017augmenting, mathur2018using}.
In addition, conventional oversampling, such as the Synthetic Minority Over-sampling Technique (SMOTE)~\cite{chawla2002smote} and the Majority Weighted Minority Oversampling Technique (MWMOTE)~\cite{barua2012mwmote} have been used to address data scarcity. 
However, since these methods are not specifically designed for data generation in HAR, they often overlook temporal dependencies and statistical properties inherent in wearable sensor data. 
This can limit their effectiveness in capturing the true underlying distribution of the data. 
As a consequence, the generated synthetic data may fail to accurately represent the complexities of real-world wearable sensor data. 

Moreover, generative adversarial network (GAN)-based methods, such as TimeGAN~\cite{yoon2019time}, have been used to generate realistic time-series data. 
However, GAN-based methods require a significant amount of labeled data for training, which is often difficult to obtain in wearable sensor-based scenarios. 
Additionally, the generated data may suffer from mode collapse or lack of diversity, limiting their effectiveness in improving the performance of HAR models.

Furthermore, biomechanical simulation-based approaches have gained popularity in generating realistic sensor data. The sensor data can be generated by capturing the real motion data and building a 3D motion model with a simulation platform. However, this method requires a substantial volume of real data to build an accurate model, which involves several steps and is time-consuming. There is a chance that the movements in the simulation model won't be as accurate as they would be in real life because of fidelity issues in the model or inaccurate input settings.

To overcome these limitations, in this paper, we propose an unsupervised statistical feature-guided diffusion model for wearable sensor-based HAR to address the challenge of costly and hard-to-annotate wearable sensor data. 
Particularly, we leverage the abundance of unlabeled wearable sensor data that can be easily obtained in real-world scenarios. 
By utilizing unsupervised learning, we generate synthetic sensor data that can enhance the performance of HAR models without relying on labeled data for training.

The diffusion model is conditioned by statistical information such as mean, standard deviation, Z-score, and skewness. 
By capturing the statistical properties of real-world wearable sensor data, it can generate synthetic data that closely resembles the characteristics of real wearable sensor data. 
Unlike traditional generative models that require class labels, our approach does not depend on labeled data, making it highly applicable to scenarios where labeled data is scarce or unavailable.
Our framework consists of two steps: the training of the unsupervised statistical feature-guided diffusion model on a large amount of unlabeled data, and the training of an independent human activity classifier using a combination of a small amount of labeled real data as well as the synthetic data generated by the pretrained diffusion model. 
This two-step process enables us to leverage the benefits of both unsupervised learning and supervised classification, leading to improved HAR performance. Moreover, since our model captures motion information from statistical features rather than labels, unlike other data synthesis methods (e.g., TimeGAN~\cite{yoon2019time}), there is no need to train separate generation models for each activity class. Data from various classes can be generated using a single trained diffusion model, significantly enhancing training efficiency.

To evaluate the effectiveness of the approach, we conduct experiments on three public openly-accessible datasets: 
MM-FIT~\cite{stromback2020mm}, PAMAP2~\cite{reiss2012introducing}, and Opportunity~\cite{chavarriaga2013opportunity}. 
We compare the method with conventional oversampling techniques, such as SMOTE~\cite{chawla2002smote} and SVM-SMOTE~\cite{nguyen2011borderline}, and a GAN-based method, TimeGAN~\cite{yoon2019time}. 
The experimental results demonstrate superiority both in terms of performance metrics for HAR and the diversity of synthetically generated data.

Our main contributions are: 
\begin{itemize}
    \item We propose an unsupervised statistical feature-guided diffusion model (SF-DM) for wearable sensor data generation in HAR, capturing the features of time-series sensor data more effectively.
    \item The diffusion model generates synthetic sensor data conditioned on statistical information, such as mean, standard deviation, Z-score, and skewness, without relying on class label information. Thus, a single trained diffusion model can be utilized to generate data from various classes.
    \item Our approach is applicable to real-world scenarios where labeled data is scarce or unavailable and requires no post-processing of the generated data; it can be utilized directly to train a HAR model.
    \item Experimental results on public openly-accessible datasets. 
    We demonstrate improved performance in HAR over conventional oversampling methods and GAN-based methods.
\end{itemize}

The rest of the paper is organized as follows. \cref{sec:relatedwork} introduces the related works. \cref{sec:proposedmethod} provides the details of the proposed diffusion model. \cref{sec:experimentalresults} presents quantitative experimental results on the three datasets. \cref{sec:discussion} discusses the potential of the proposed method and available applications in HAR. Finally, \cref{sec:conclusion} concludes the paper.

\section{Related Work}
\label{sec:relatedwork}
\subsection{Wearable Sensor-based Human Activity Recognition}

Sensor-based systems have dominated the applications of monitoring our daily activities, given the privacy concerns associated with placing cameras in our personal space~\cite{gheid2016novel,qi2018hybrid,zhang2019collective} and highlighting the importance of sensor-based HAR. In Park et al.'s work~\cite{park2023multicnn}, they proposed a MultiCNN-FilterLSTM model to provide a resource-efficient method for HAR on smart devices. \cite{dahou2023mlcnnwav} developed a discrete wavelet transform method coupled with a multilayer residual convolutional neural network (MLCNNwav) for sensor-based HAR, increasing generalization and minimizing processing complexity. Ferrari et al.~\cite{ferrari2023deep} proposed a personalization technique combined with machine learning to improve the generalization ability of the model. 

Given the effectiveness and widespread adoption of Transformer architecture~\cite{vaswani2017attention}, it is also introduced in sensor-based HAR. Dirgova et al.~\cite{dirgova2022wearable} adapted the transformer for time-series analysis and achieved a classification accuracy of 99.2\% on a public smartphone motion sensor dataset that covers a wide range of activities. \cite{xiao2022two} proposed a self-attention-based Two-stream Transformer Network (TTN) to capture the temporal and spatial information, respectively. Pramanik et al.~\cite{pramanik2023transformer} presented a deep reverse transformer-based attention mechanism, exploiting a top-down feature fusion. The reverse attention regularizes the attention modules and adjusts the learning rate adaptively. Feng et al.~\cite{feng2024atfa} introduced the Adversarial Time-Frequency Attention (ATFA) framework to effectively address data heterogeneity issues caused by increased sensor use and diverse user contexts for sensor-based HAR.


Due to the scarcity of labeled wearable sensor datasets, HAR systems are also trained with unlabeled data to improve performance and robustness. Semi-supervised and unsupervised learning are introduced to tackle the challenge. In Balabka et al.'s work~\cite{balabka2019semi}, they trained the adversarial autoencoder with unlabeled data and validated the model with a small amount of labeled data. Oh et al.~\cite{oh2021study} combined the existing active learning with semi-supervised learning and achieved outstanding performance with less labeled data. For unsupervised learning, Sheng et al.~\cite{sheng2020unsupervised} used K-means clustering and autoencoder to accurately classify fully unlabeled wearable sensor data. LLMIE-UHAR~\cite{gao2024unsupervised} leveraged large language models (LLMs) and Iterative Evolution (IE) to achieve an unsupervised way for HAR.

To address the scarcity of annotated data and accommodate the diverse real-world settings in which HAR is conducted (sensor modalities, downstream tasks, etc.), transfer learning and contrastive learning are employed. RecycleML~\cite{xing2018enabling} used cross-modal transfer to accelerate the learning of knowledge between edge devices using unlabeled data. Banos et al.~\cite{banos2021opportunistic} introduced transfer learning to train new, unseen, or target sensor systems opportunistically by leveraging existing or source sensor systems. The approach employs system identification techniques to acquire a mapping function, enabling automatic translation of signals from the source sensor domain to the target sensor domain, and vice versa. COCOA~\cite{deldari2022cocoa} explored sensor-based cross-modal contractive learning, achieving quality representations from multisensory data through cross-correlation computations across various data modalities and mitigating the similarity among irrelevant instances. FOCAL~\cite{liu2024focal} proposed a novel contrastive learning framework for multimodal time-series sensing signals by introducing orthogonality constraint, which outperforms the SOTA in downstream tasks including HAR.

\subsection{Sensor Data Generation for Human Activity Recognition}
The scarcity of labeled data in wearable sensor applications presents a significant challenge, prompting the exploration of synthetic data generation as a viable solution. While extensive research has been conducted in computer vision and natural language processing domains, the field of wearable sensor data generation remains relatively understudied, necessitating tailored methods to accommodate the unique characteristics of such data. Three representative approaches have emerged to address the challenge of generating wearable sensor data: 1) traditional oversampling techniques, 2) virtual IMU data generation from alternative modalities, and 3) generating sensor data using generative adversarial networks (GAN).

Traditionally, data augmentation techniques have been employed to address data scarcity issues in various domains \cite{kim2021label, jeong2021sensor}. While conventional approaches in computer vision often rely on simple affine transformations for data augmentation, such as translation and rotation \cite{shorten2019survey, maharana2022review}, the temporal nature of accelerometer signals in sensor-based HAR necessitates alternative strategies. Signal-processing methods, including jittering, scaling, and random sampling, have been proposed to augment accelerometer signals, effectively enhancing the diversity of training datasets \cite{um2017data, ohashi2017augmenting, mathur2018using, cheng2023learning}. In addition to the signal processing methods, traditional oversampling techniques like SMOTE and MWMOTE have been employed to mitigate data scarcity. However, these methods, while employed in sensor-based HAR, may overlook temporal dependencies and statistical properties inherent in wearable sensor data, limiting their effectiveness in capturing the complexities of real-world data distributions \cite{alharbi2022comparing}.

Secondly, efforts to bridge the gap between video and Inertial Measurement Unit (IMU) data have gained traction, with recent works focusing on translating video data into IMU representations \cite{rey2019let, kwon2020imutube, santhalingam2023synthetic}. These endeavors leverage generative methods \cite{rey2019let, fortes2021translating, santhalingam2023synthetic} and trajectory-based approaches \cite{kwon2020imutube, xiao2021deep, banos2012kinect} to extract IMU data from videos, expanding the applicability of sensor-based HAR beyond traditional IMU-equipped scenarios. Generative methods employ machine learning to derive IMU data directly from videos \cite{rey2019let, fortes2021translating}, whereas trajectory-based methods estimate joint orientations from 3D joint positions extracted from videos \cite{kwon2020imutube, xiao2021deep}. These approaches have demonstrated success in generating synthetic IMU data for human activity recognition tasks, enhancing model performance and generalizability across diverse scenarios. Despite the efficacy of systems like IMUTube, challenges remain, particularly concerning the quality of input videos. The limitations of vision-based systems are evident when videos exhibit camera egomotion or include irrelevant scenes, requiring meticulous video selection. 

Thirdly, GAN-based methods have shown promise in generating realistic time-series data, combining unsupervised and supervised training approaches \cite{yao2018sensegan, wang2018sensorygans, yoon2019time}. SenseGAN \cite{yao2018sensegan} and SensoryGANs \cite{wang2018sensorygans}, for instance, have introduced frameworks for generating synthetic sensor data, effectively improving human activity recognition in resource-limited environments. TimeGAN \cite{yoon2019time} and ActivityGAN \cite{li2020activitygan} have demonstrated superior performance in maintaining temporal dynamics and augmenting sensor-based HAR datasets, respectively. Furthermore, Balancing Sensor Data Generative Adversarial Networks (BSDGAN) \cite{hu2023bsdgan} address imbalanced datasets in HAR, effectively enhancing recognition accuracy for activity recognition models. A time-series GAN (TS-GAN) \cite{yang2023ts} based on LSTM networks for augmenting sensor-based health data was proposed to improve the performance of deep learning-based classification models. TS-GAN utilized an LSTM-based generator and discriminator, incorporating a sequential-squeeze-and-excitation module and gradient penalty from Wasserstein GANs for stability. 
However, these GAN-based methods demand a substantial amount of labeled data for training, a challenge in wearable sensor-based scenarios. Mode collapse and a lack of diversity in generated data are additional concerns that may limit their efficacy in improving HAR model performance. 

Lastly, the biomechanical simulation-based approaches are attracting increasing interest from researchers. Jiang et al.~\cite{jiang2021model} utilized OpenSim, an open-source software system for biomechanical modeling, simulation, and analysis, to simulate individuals with various physiological characteristics performed movements to augment the IMU dataset. Uhlenberg et al.~\cite{uhlenberg2023co} generated synthetic accelerations as well as angular velocities with a simulation framework to enable a comprehensive analysis of gait events. Tang et al.~\cite{tang2024synthetic} utilized the simulation platform OpenSim and forward kinematic methods to generate a substantial volume of synthetic IMU data for fall detection. However, the implementation of the method is complicated and time-consuming, it involves several steps. Firstly, recordings of multi-view actions from participants wearing IMU sensors need to be collected, followed by a pose estimation. Then calculate coordinates so that the simulation platform can accurately represent and simulate physical interactions and movements. It has the potential for inaccuracies or discrepancies between simulated and real-world movements due to the limitations in the simulation model's fidelity or inaccuracies in input parameters.

\subsection{Diffusion Probabilistic Models}

The idea of the diffusion model is inspired by non-equilibrium statistical physics, which is to gradually eliminate structure in a data distribution using an iterative forward diffusion approach. 
Then, a reverse diffusion process that reinstates structure in the data is learned, producing an adaptable and manageable generative model~\cite{SohlDickstein2015DeepUL}. 
Recently, Diffusion Probabilistic Models (DM) beat GAN~\cite{Goodfellow2014GAN} and have achieved state-of-the-art results in image synthetic by ensuring high quality and diversity~\cite{Ho2020DDPM,nichol2021improved,Dhariwal2021DMbeatGANs,Rombach2021HighResolutionIS,Ho2022VideoDM,Gu2021VectorQD}. 
The authors in~\cite{Ho2020DDPM} represent the diffusion process (forward process) as a Markov chain that transforms the original data distribution into a Gaussian distribution and the reverse process (denoising process) learns to generate samples by removing noise step by step with a deep learning model. 
The U-Net~\cite{ronneberger2015u} architecture is considered to be a powerful deep learning model for denoising~\cite{Ho2020DDPM,Ho2022VideoDM,JolicoeurMartineau2020AdversarialSM}. 
To further improve the denoising performance, a stack of residual layers and attention mechanisms are introduced to U-Net-like models~\cite{Rombach2021HighResolutionIS,song2021scorebased,kim2022flame}. The application of the diffusion model often necessitates thousands of computation steps to obtain a high-quality new sample, thereby imposing significant limitations on its practical usability. 
Methods to enhance the sampling speed of a diffusion model are discretization optimization~\cite{dockhorn2022genie}, utilizing a non-Markovian process~\cite{song2020denoising} and partial sampling~\cite{song2020denoising}.

In addition to image synthesis, diffusion models have been applied to other tasks including image in-painting~\cite{lugmayr2022repaint}, 3D shape generation~\cite{zhou20213d}, text generation~\cite{gong2022diffuseq}, audio synthesis~\cite{kong2020diffwave}, molecular conformation generation~\cite{xu2022geodiff}, etc. 
Shao et al.~\cite{shao2023study} utilize a diffusion model with redesigned UNet to generate synthetic sensor data for HAR by incorporating the label information. Huang et al.~\cite{huang2023diffar} propose an adaptive conditional diffusion model to improve the HAR performance based on channel state information (CSI) by augmenting CSI based on visualized spectrum information. In contrast, we propose the first unsupervised statistical feature-guided diffusion model for sensor data generation in HAR, which operates without the need for labeled information and can be directly applied to raw sensor data.

\section{Method}
\label{sec:proposedmethod}

\begin{figure}[!t]
\includegraphics[width=0.7\columnwidth]{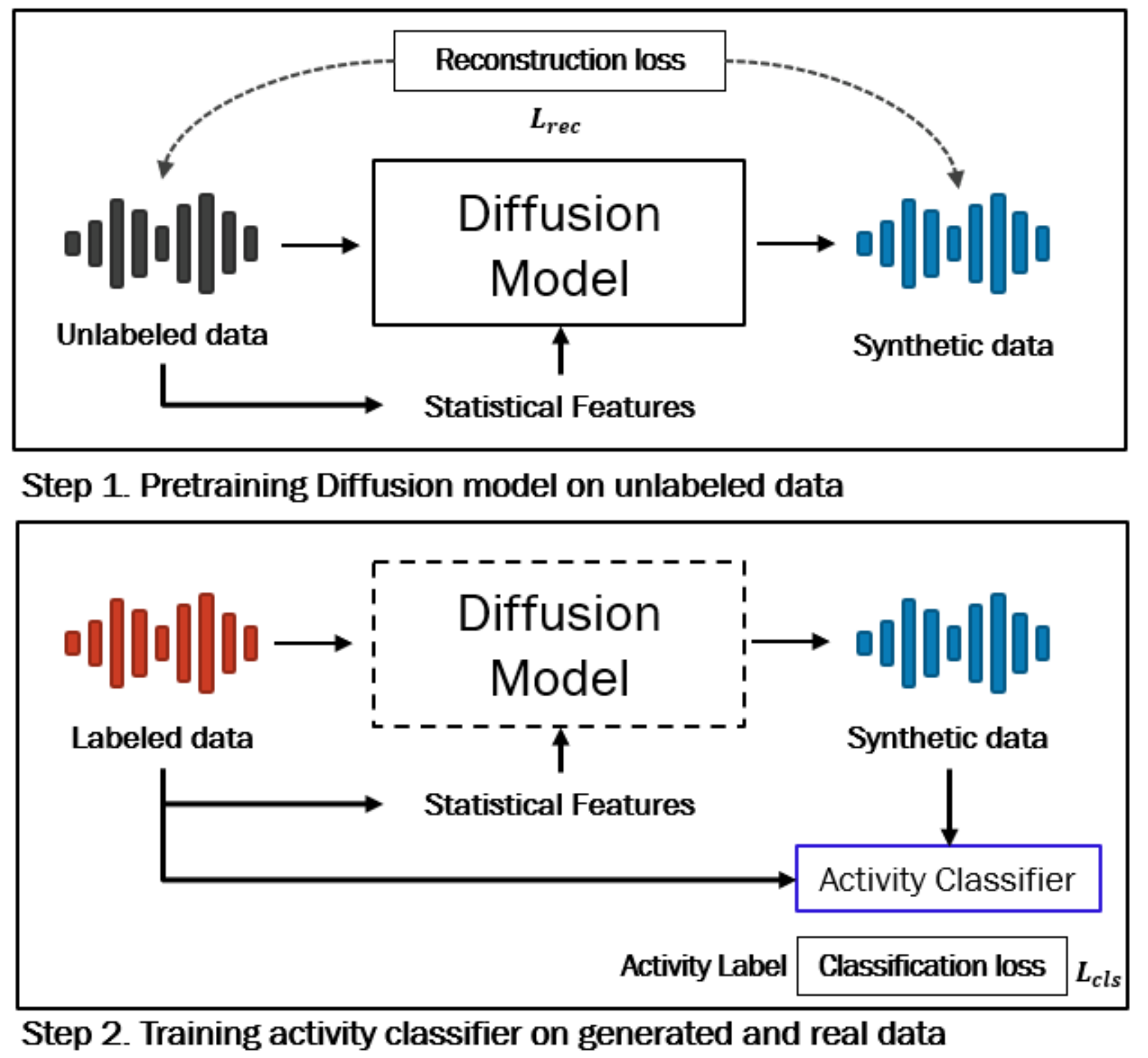}
\caption{Overview of the statistical feature-guided diffusion model (SF-DM). 
Step 1: pretrain SF-DM with unlabeled real data; Step 2: train the HAR classifier with synthetic data generated by SF-DM and finetune the classifier with real data.
}
\label{fig:teaser}
\end{figure}

We propose to improve the performance of HAR models through a two-step process, as depicted in \cref{fig:teaser}.
First, we pretrain the diffusion model using real unlabeled sensor data with statistical features. Second, we train the HAR classifier using synthetic sensor data generated by the well-trained diffusion model (\textbf{SF-DM}), followed by fine-tuning the classifier with real data. 

\subsection{Structure of SF-DM}
\begin{figure}[!t]
 \centering
\includegraphics[width=0.9\linewidth]{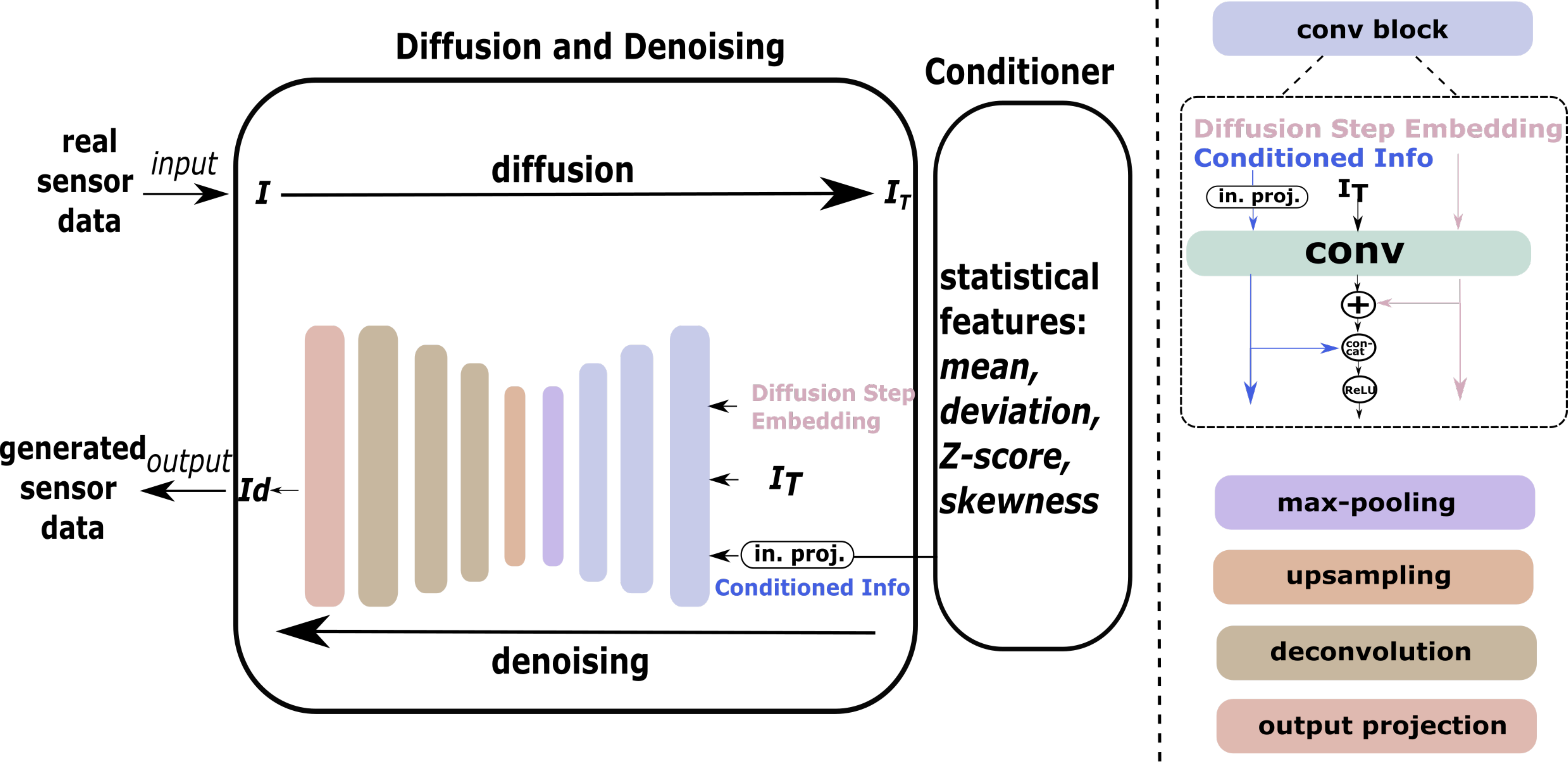}
\caption{The architecture of the diffusion model. The diffusion model consists of the diffusion and denoising modules. The brief architecture of the denoising module is depicted on the right side. The denoising module receives statistical features as a conditioner and the architecture of the denoising module is based on the U-Net architecture.
}
\label{fig:diffusion}
\end{figure}

The unsupervised statistical feature-guided diffusion model consists of two main components: the diffusion and denoising model, as well as the conditioner. As depicted in \cref{fig:diffusion}, we employ an encoder-decoder framework for \textbf{SF-DM}.

In the diffusion stage, we first randomly sample a time step $t$ with the range $(0, T]$ where $T$ is the maximum diffusion step. The level of noise varies according to the current diffusion step $t$. With a larger step, the noise becomes more pronounced. The noisy data is created by a weighted sum of noise and data (see~\cref{eq:noise_data}).

\begin{figure}[!t]
\includegraphics[height=15cm]{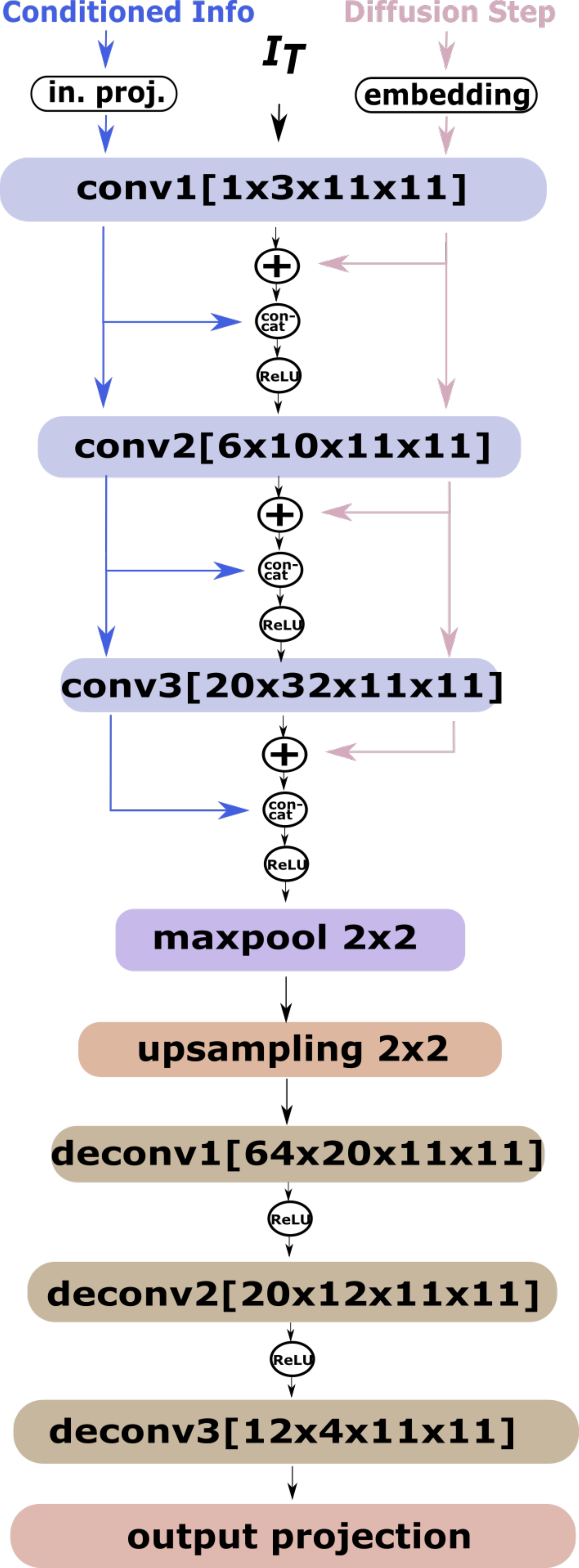}
\caption{Unsupervised statistical feature-guided diffusion model. $I_T$ refers to the sensor data after diffusion step $T$.
}
\label{fig:diffusionarchitecture}
\end{figure}

The denoising model consists of an encoder-decoder framework. The encoder of \textbf{SF-DM}, which learns features from the input, includes three convolutional blocks and a max-pooling layer. The convolutional block has three different inputs: 
statistical features from the conditioner, noisy data \textbf{$I_T$} from diffusion step \textit{T}, and diffusion step embedding. 
Before feeding the statistical features to the convolutional layer, we project them to match the shape of the noisy data. 
The output of diffusion step embedding from the convolutional layer is added to that of \textbf{$I_T$} and concatenated with the output of statistical features from the convolutional layer for providing additional information. 
We employ convolutional layers with a kernel size of 9$\times$9 and a max-pooling layer with a kernel size of 2$\times$2 and stride 2.
The decoder incorporates an upsampling layer that restores the resolution to match that of the previous layer and deconvolutional layers~\cite{zeiler2010deconvolutional} (with a kernel size of 9$\times$9) which disseminates the information contained in one data point across multiple data points. An output projection layer is followed to match the dimension of the output from the diffusion model with the input real data. A detailed structure of the proposed model is shown in~\cref{fig:diffusionarchitecture}.

\subsection{Statistical features selection}
To guide the generation of synthetic sensor data, we leverage statistical features extracted by the conditioner from real data. Compared with only using labels, these statistical features provide richer information about the activity.
The features include mean, standard deviation, Z-score $\left(\frac{x-\mu}{\sigma}\right)$, where $x$ is an observed value, $\mu$ represents the mean of all values, and $\sigma$ indicates the standard deviation of the sample), and skewness ($\gamma = \mathbb{E}[\left(\frac{x-\mu}{\sigma}\right)^3]$), which measures the asymmetry of the probability distribution. 
Importantly, label information is not required, which means the \textbf{SF-DM} can be widely applied to unsupervised training with unlabeled datasets. 
We concatenate three features before input to the diffusion model.

\subsection{HAR with SF-DM}

\begin{figure}[!t]
 \centering
\includegraphics[width=0.3\columnwidth, height=5.5cm]{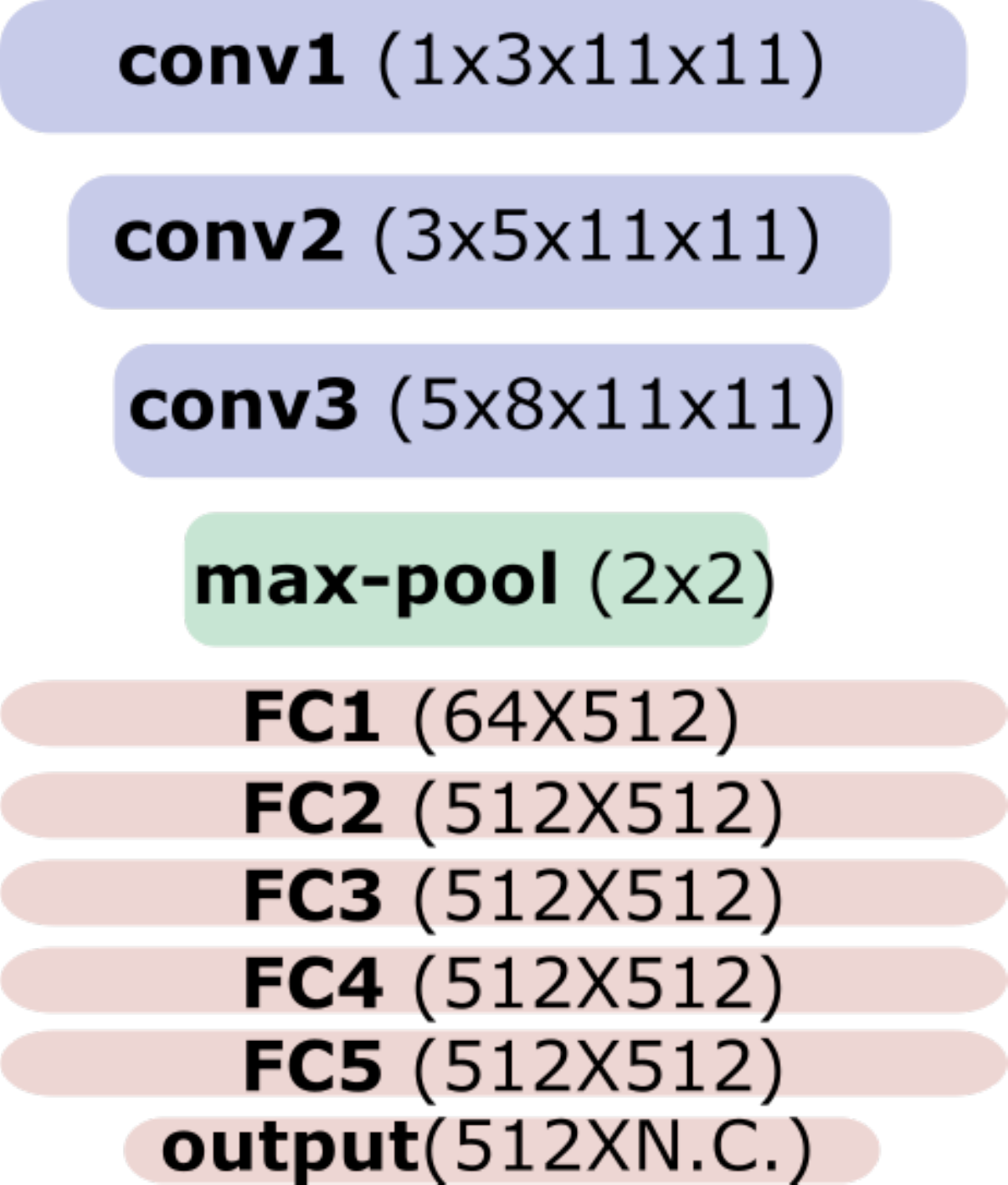}
\caption{The architecture of HAR classifier.
}
\label{fig:classifier}
\end{figure}

In this work, we employ a simple convolutional neural network (CNN) for HAR, as illustrated in \cref{fig:classifier}.
The architecture comprises three convolutional layers (\textbf{conv1}, \textbf{conv2}, \textbf{conv3}) with a stride of 2, each layer followed by a ReLU activation function, along with a max-pooling layer with dilation 1. Additionally, there are five fully connected layers (\textbf{FC1}, \textbf{FC2}, \textbf{FC3}, \textbf{FC4}, \textbf{FC5}), each followed by a ReLU activation function, and an output layer. 
We use the same structure for both pretraining and fine-tuning.

As depicted in~\cref{fig:teaser}, the training procedure of the HAR classifier with \textbf{SF-DM} includes two key steps:
\begin{enumerate}
    \item \textbf{SF-DM} training: As \textbf{SF-DM} does not require label information, we first train \textbf{SF-DM} on a large volume of unlabeled real data
    \item HAR classifier training: We initially pretrain the HAR classifier using synthetic sensor data generated by the pretrained \textbf{SF-DM} and then finetune the classifier with real data (\cref{alg:sfdm})
\end{enumerate}
For \textbf{SF-DM} training, in each training iteration, we sample a batch of real data and calculate the corresponding statistical features, including mean $\mu$, standard deviation $\sigma$, Z-score $z$, and skewness $\gamma$, which have the same length as the input data sequences, and concatenate them as $f$. We then generate noisy data $\tilde{x}$ from the sampled real sensor data as follows:
\begin{equation}
\label{eq:noise_data}
    \tilde{x} = x \times \sqrt{\beta[t]} + \epsilon \times \sqrt{1-\beta[t]}
\end{equation}
In \cref{eq:noise_data}, $x$ denotes the real sensor data, $\beta$ indicates the noise level, $t$ represents the diffusion step, and $\varepsilon$ denotes the random noise that has the same shape as the input real data. The $\tilde{X}$ together with statistical features are then fed into the \textbf{SF-DM}.

The diffusion model generates synthetic data by removing noise, and thus, the $SF-DM$ is trained by minimizing a reconstruction loss between the original real wearable sensor data and the generated data.
\begin{equation}
\label{eq:mae}
    L_{\mbox{\tiny{rec}}} (x, \tilde{x}, f;\theta_E) =\frac{\sum_{l=1}^{n} |D(\tilde{x}_l, f_l) - x_l|}{n}
\end{equation}
In \cref{eq:mae}, $x$ represents the unlabeled real sensor data, $\tilde{x}$ is the input noisy data, $f$ denotes the statistical features including mean, standard deviation, and Z-score of $x$, $D$ indicates the diffusion model with the decoder and encoder, and $n$ is the number of data samples.
In this procedure, the \textbf{SF-DM} can be trained in an unsupervised statistical feature-guided way. 

\begin{algorithm}
	\caption{Training procedure for unsupervised statistical feature-guided diffusion model (SF-DM) and human activity classifier with generated data}
 \label{alg:sfdm}
	\begin{algorithmic}[1]
            \STATE \textbf{Step 1: \textit{SF-DM training}} 
            \REQUIRE Batch size $m$, Adam hyperparameter $\eta_E$
    	\FOR {Number of training iterations for Step 1}
                \STATE Sample a batch $x$ from the training dataset $X$.
                \STATE Calculate the statistical features: $\mu$, $\sigma$, $Z$, $\gamma$
                \STATE $f \gets concat(\mu, \sigma, Z, \gamma)$
                \STATE Generate noise data  $\epsilon \sim N(0,1)$ from real data with the noise scale $\beta$ and the diffusion step $t$:\\
                $\tilde{x} \gets x\sqrt{\beta[t]} + \epsilon\sqrt{1-\beta[t]}$
                \STATE $\theta_E \gets \theta_E - \eta_E \nabla_{\theta_E} L_{\mbox{\tiny{rec}}}(x,\tilde{x},f;\theta_E)$ 
                \hfill$\triangleright$\cref{eq:mae}
            \ENDFOR
            \STATE \hrulefill
            \STATE \textbf{Step 2: \textit{HAR classifier training}}
            \REQUIRE Batch size $m$, Adam hyperparameter $\eta_C$
            \FOR {Number of training iterations}
                \STATE Sample a batch $(x,y)$ from the training dataset $X$ and corresponding activity label $Y$.
                \STATE Calculate the statistical features: $\mu$, $\sigma$, $Z$, $\gamma$
                \STATE $f \gets concat(\mu, \sigma, Z, \gamma)$
                \STATE Make a random noise vector: $\omega \sim N(0,1)$
                \STATE $\theta_C \gets \theta_C - \eta_C \nabla_{\theta_C} L_{\mbox{\tiny{syn}}}(\omega, f, y;\theta_C)$
                \hfill$\triangleright$\cref{eq:crossEntropy} 
            \ENDFOR
            \FOR {Number of training iterations}
                \STATE Sample a batch $(x,y)$ from the training dataset $X$ and corresponding activity label $Y$.
                \STATE $\theta_C \gets \theta_C - \eta_C \nabla_{\theta_C} L_{\mbox{\tiny{real}}}(x,y;\theta_C)$
                \hfill$\triangleright$\cref{eq:realcrossEntropy} 
            \ENDFOR
	\end{algorithmic} 
\end{algorithm} 

Next, the HAR classifier is pretrained on synthetic data from SF-DM. 
From each batch of real data, we first calculate the statistical features and initialize random noise with a shape that matches the input of the \textbf{SF-DM}. 
Combining random noise and statistical features, SF-DM is able to produce synthetic sensor data sequences and patterns that resemble real data generated by the same type of sensor. 
The HAR classifier is trained with the synthetic sensor data and the corresponding activity label.
\begin{equation}\label{eq:crossEntropy}
    L_{\mbox{\tiny{syn}}} (\omega, f,y;\theta_C) =-{\sum_{l=1}^{n_{c}} y_l \log C(E(\omega_l, f_l)))}
\end{equation}
where $x$ represents the input data, $y$ is the corresponding class label, $E$ denotes the pretrained diffusion model, $C$ is the activity classifier, $\omega$ is the random noise for an input of the diffusion model, $f$ denotes the statistical features of $x$, and $n_c$ indicates the number of classes. 

To further improve the HAR performance, we fine-tune the HAR classifier with real sensor data.
\begin{equation}\label{eq:realcrossEntropy}
    L_{\mbox{\tiny{real}}} (x,y;\theta_C) =-{\sum_{l=1}^{n_{c}} y_l \log C(x_l)}
\end{equation}

In summary, the unsupervised statistical feature-guided diffusion model (\textbf{SF-DM}) and HAR classifier are trained in a two-step process to improve wearable sensor-based human activity recognition. 
The \textbf{SF-DM} is trained in an unsupervised statistical feature-guided way, while the HAR classifier is pretrained with synthetic sensor data and then fine-tuned with real sensor data. 
The training details for the unsupervised statistical feature-guided diffusion model (\textbf{SF-DM}) and HAR classifier are provided in \cref{alg:sfdm}.

\section{Experimental Results}\label{sec:experimentalresults}

\begin{table}[!t]
  \caption{Comparison of \textit{MM-Fit}, \textit{PAMAP2}, and \textit{Opportunity (locomotion)} datasets  }
  \begin{center}
  \begin{tabular}{c|c|c|c}
    \hline
      & \textit{MM-Fit} &\textit{PAMAP2}   &  \textit{Opportunity} 
     \\
    \hline
    Num. of activities  & 10  & 18 & 4\\ \hline
    Num. of subjects  & 21  & 9 & 4\\ \hline
    Sensor  & Accelerometer & Accelerometer & Accelerometer\\ \hline
    Frequency(Hz)  & 100 & 100 & 30\\ \hline
    Total Duration (Sec.) & 48540 & 27248.27 & 20240.4 \\\hline 
    Position  & wrist & wrist & knee\\ \hline
\end{tabular}
\end{center}
\label{tab:datasets_details}
\end{table}

\subsection{Datasets} 
In this section, we conducted comprehensive evaluations of \textbf{SF-DM} on three different public datasets: \textit{MM-Fit}~\cite{stromback2020mm}, \textit{PAMAP2}~\cite{reiss2012introducing}, and \textit{Opportunity}~\cite{chavarriaga2013opportunity}. The detailed information of the datasets is summarized in \cref{tab:datasets_details}. 

\textit{MM-Fit} comprises wearable sensor data collected from diverse time-synchronized devices during ten full-body exercises by multiple subjects.
The data is collected from different devices, including depth cameras, smartphones, smartwatches, and earbuds, capturing accelerometer, gyroscope, and magnetometer modalities. 
For our experiment, we utilized the accelerometer data of the smartwatch worn on the left hand with a sampling frequency of 100 Hz. 

\textit{PAMAP2} contains data from nine subjects (one female, eight right-handed) wearing three inertial measurement units and a heart rate monitor while engaging in 18 different physical activities. 
In our experiment, we used the protocol set. We chose the accelerometer sensor data from the IMU sensor placed on the dominant hand, sampled at 100 Hz. Subject 109 was excluded as they performed only a few motions. We restricted our analysis to those motions executed by all the selected subjects (ironing, lying, sitting, standing, walking, ascending stairs, descending stairs, vacuum cleaning, and non-activity). Data from subjects 101-106 were used for training, data from subject 107 was chosen for validation, and subject 108's data was used for testing.

\textit{Opportunity} dataset captures on-body sensor data during naturalistic human activities, categorized into \textbf{Drill} (sequential pre-defined activities) and \textbf{ADL} (a high-level task with flexible atomic activity sequence) sessions. 
A challenge of the \textit{Opportunity} dataset is that it comprises recordings of 4 participants using solely on-body sensors, and five unsegmented recordings for each subject are provided. 
We only consider the locomotion and data from the accelerometer \textit{RKN\_} (placed on the right leg below the knee). The reason is that there is a large amount of missing data from the sensor that was placed on the wrist, and we chose the locomotion task as the target activity for classification. During the implementation, we followed the recommendations of the paper that proposed the sets. Only data from subjects 1, 2, and 3 were used. We kept ADL4 and ADL5 from subjects 2 and 3 for testing, ADL5 from subject 1 for validation, and the remaining for training.

To ensure fairness, we conducted a five-fold cross-validation, separating data based on subjects. For example, for \textit{MM-Fit} dataset, training data include subjects '01', '02', '03', '04', '06', '07', '08', '14', '15', '16', '17', and '18', with subject '19' for validation, and testing conducted on subjects '09', '10', and '11'.
The window size (number of data points per window) for \textit{MM-Fit}, \textit{PAMAP2} and \textit{Opportunity} is 400, 200, and 200, respectively. 
The label of window data is assigned based on the majority class, determined by the class that has the highest number of data points within the window.
Across all datasets, we use the Euclidean norm ($\sqrt{x^2+y^2+z^2}$) of the accelerometer data collected from the three axes (\textit{x}, \textit{y}, \textit{z}) as input.

\subsection{Implementation Details}
All experiments were conducted in Python using the PyTorch framework on a Linux system with a Tesla P100 GPU.
We chose the Adam optimizer~\cite{kingma2014adam} and initialized the learning rate with a value of 0.0002. 
The step size of the forward diffusion process is controlled by a variance schedule $\begin{Bmatrix} \beta_t \in (0.0001, 0.05) \end{Bmatrix}$ where $t$ ranges from 1 to \textbf{T}. 
The maximum diffusion step is set to be 50.
The batch size was 128, and training spanned 200 epochs with early stopping using a patience of 20 epochs to counter overfitting.

\subsection{Comparison Baseline}

\begin{figure}[!t]
    \centering
    \includegraphics[width=0.6\linewidth]{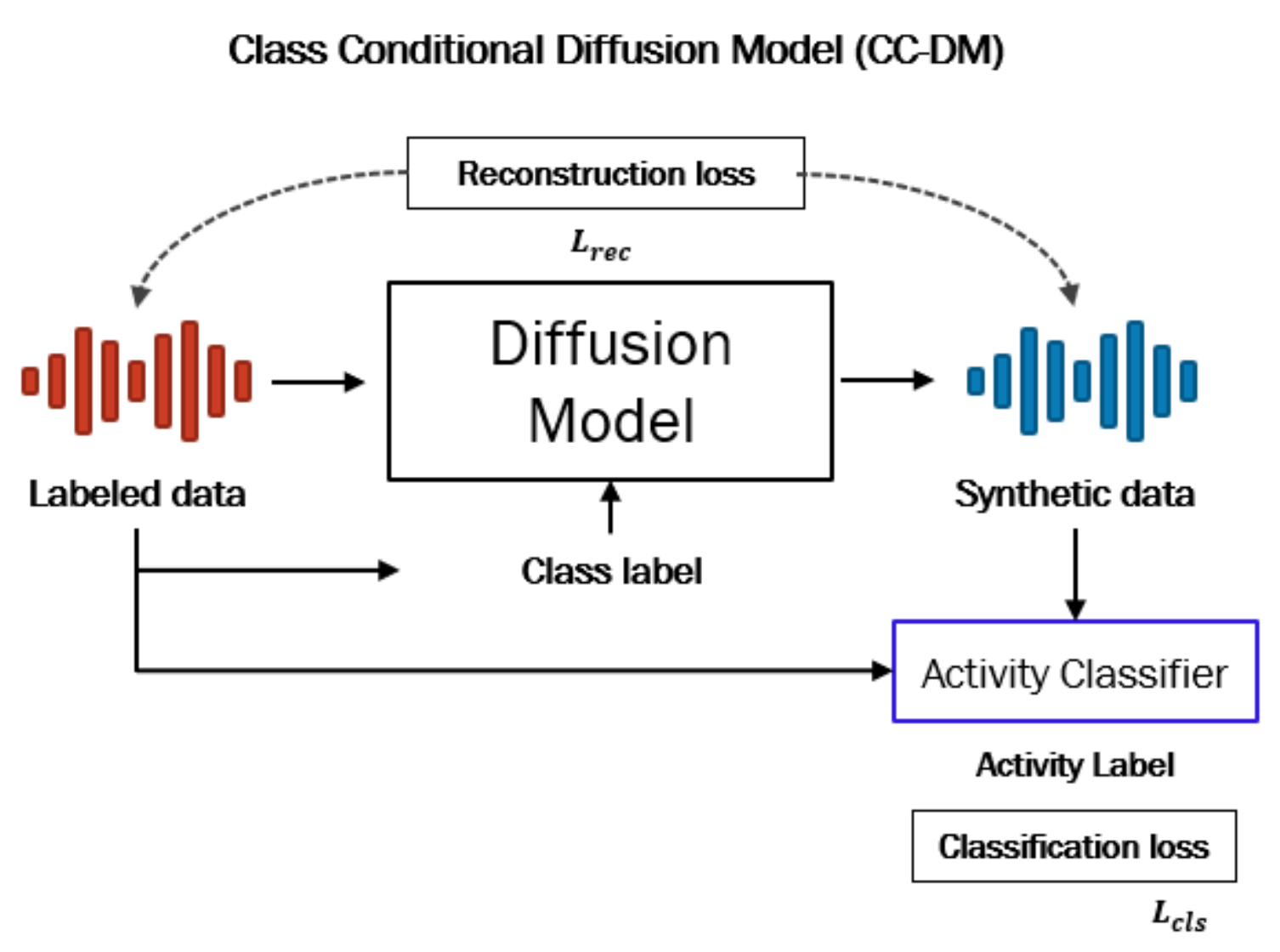}
    \caption{HAR with Class Conditional Diffusion Model.}\label{fig:ccdm}
\end{figure}

To evaluate the classification performance, we compare our model with conventional oversampling techniques (SMOTE~\cite{chawla2002smote} and SVM-SMOTE~\cite{nguyen2011borderline,balaha2023comprehensive}) and a GAN-based approach (TimeGAN~\cite{yoon2019time}). It is crucial to mention that existing GAN-based models \cite{yao2018sensegan,wang2018sensorygans,li2020activitygan,hu2023bsdgan,yang2023ts} have demonstrated limitations of diversity on generated data and pose significant challenges in the training stage due to mode collapse and unstable convergence. Only TimeGAN could be trained stably and successfully generate wearable sensor data for HAR. Hence, we exclusively selected TimeGAN as a representative GAN-based model for comparison. Furthermore, we introduce a class conditional diffusion model (\textbf{CC-DM}, see~\cref{fig:ccdm}), which is solely guided by label information. Given that the structure of our proposed diffusion model employs an encoder-decoder framework, similar to the architecture of U-Net and conditioned only on label information, we consider it comparable to the method presented in Shao et al.'s work \cite{shao2023study}.

To ensure fairness, all models followed the same two-step training procedure: 
(1) pretraining the data generation model, 
(2) training the HAR classifier on synthetic data followed by fine-tuning with 20\% of real data. It is worth mentioning that we did not train separate models for each class with SF-DM. The concept is to generate IMU data based on specific statistical information rather than using labeled data. In contrast, for TimeGAN, we trained a separate model for each class in all three datasets.

We repeated each experiment ten times and averaged results for classification accuracy and Macro F1 score.

To effectively demonstrate the impact of \textbf{SF-DM} on enhancing the performance of HAR, we additionally train the diffusion model using varying quantities of real data, considering scenarios where data is scarce. We conduct a comparison of the Macro F1 score between two models: one trained solely on a specific quantity of real data (baseline), and another model that undergoes pre-training on synthetic sensor data from \textbf{SF-DM} (also trained on the same amount of real data) followed by fine-tuning using the same real dataset (\textbf{SF-DM[Corresp.P]}). 
We train \textbf{SF-DM} on the full dataset while having the HAR classifier trained on a certain proportion of the real data (the proportion of real data used: [0.2, 0.3, 0.4, 0.5, 1]) - \textbf{SF-DM[Proportion:1]}. 
An ablation study is conducted on \textbf{CC-DM}. We train \textbf{CC-DMs} with proportional real data and evaluate their classification performance. The evaluation of generative models is subjective. In the assessment of the diffusion model, we use accuracy and Macro F1 score on the HAR task.

\begin{figure}[!t]
    \begin{center}
    \subfigure[\textit{MM-Fit} dataset]{\label{fig:testsample_mmfit}
        \includegraphics[width=0.5\columnwidth, height=2.5cm]{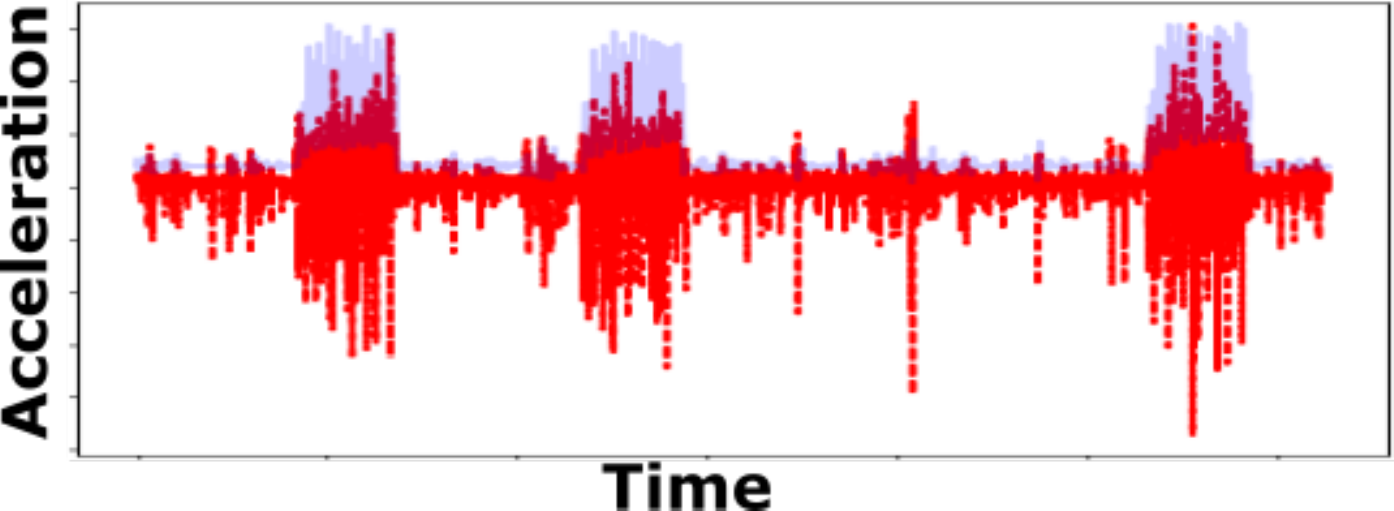}} 
    \end{center}
    \subfigure[Squats]{\label{fig:testsample_squats}
        \includegraphics[width=0.48\columnwidth, height=1.15cm]{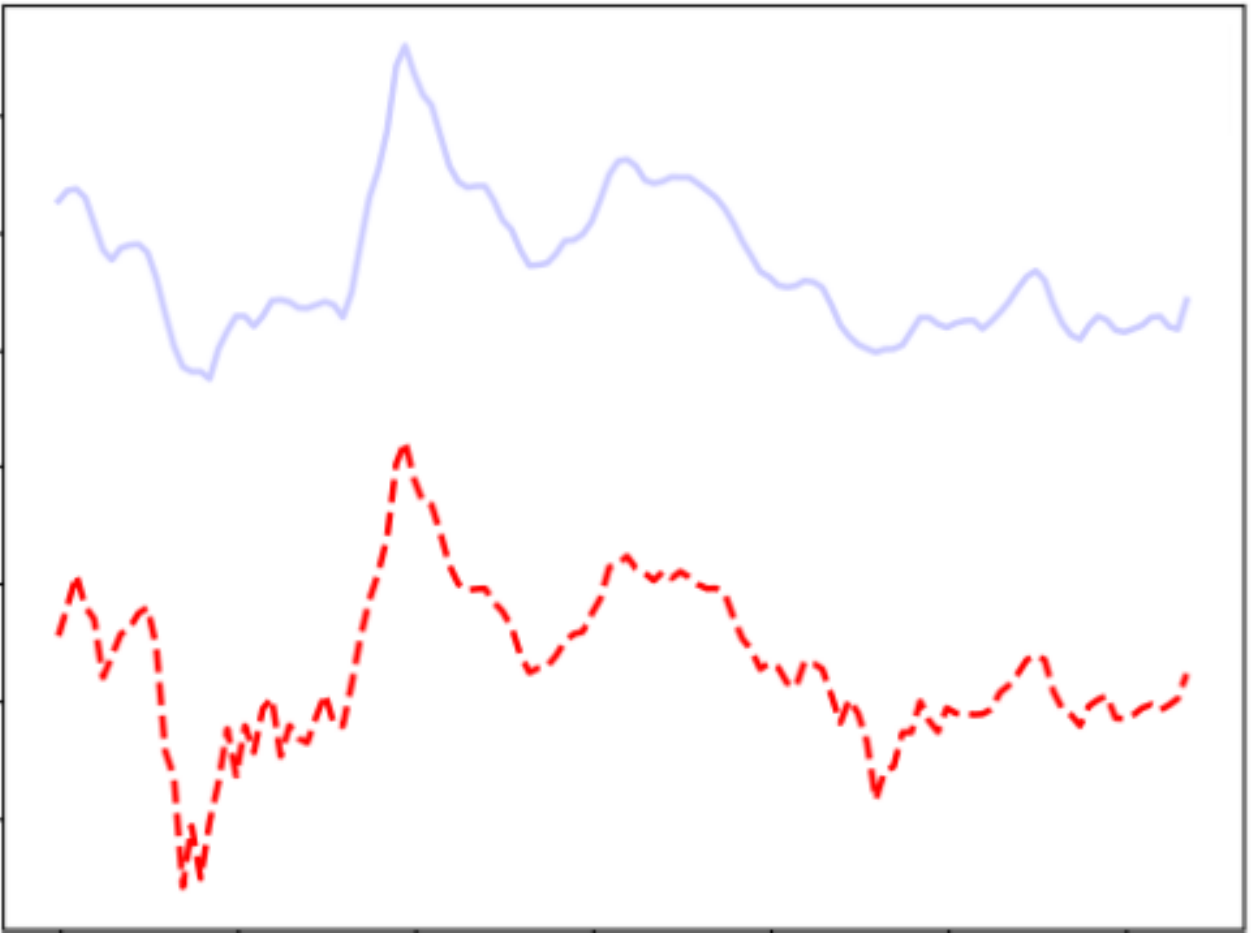}} 
    \subfigure[Lunges]{\label{fig:testsample_lunges}
        \includegraphics[width=0.48\columnwidth,height=1.15cm]{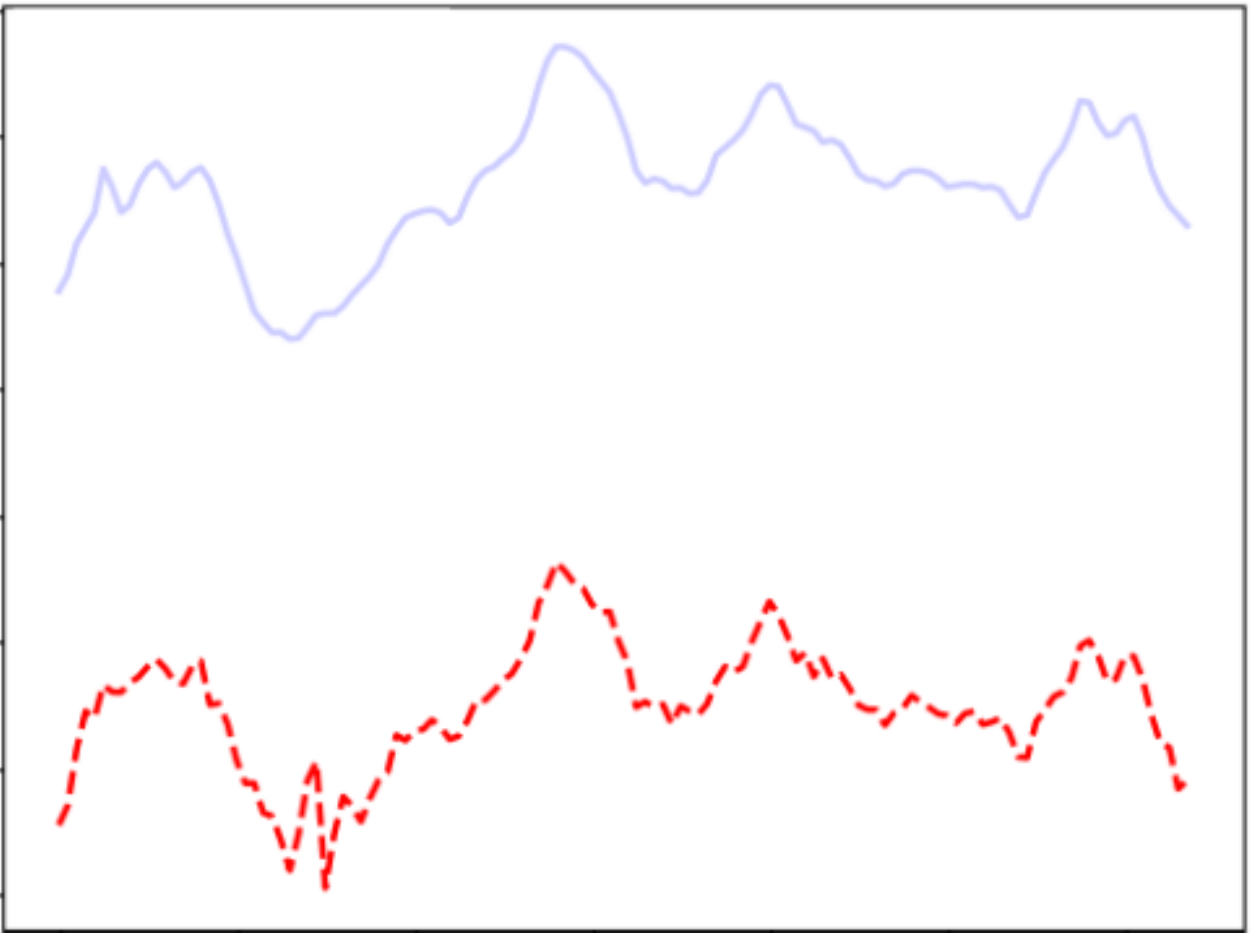}} 

    \subfigure[Bicep curls]{\label{fig:testsample_bicep_curls}
        \includegraphics[width=0.48\columnwidth, height=1.15cm]{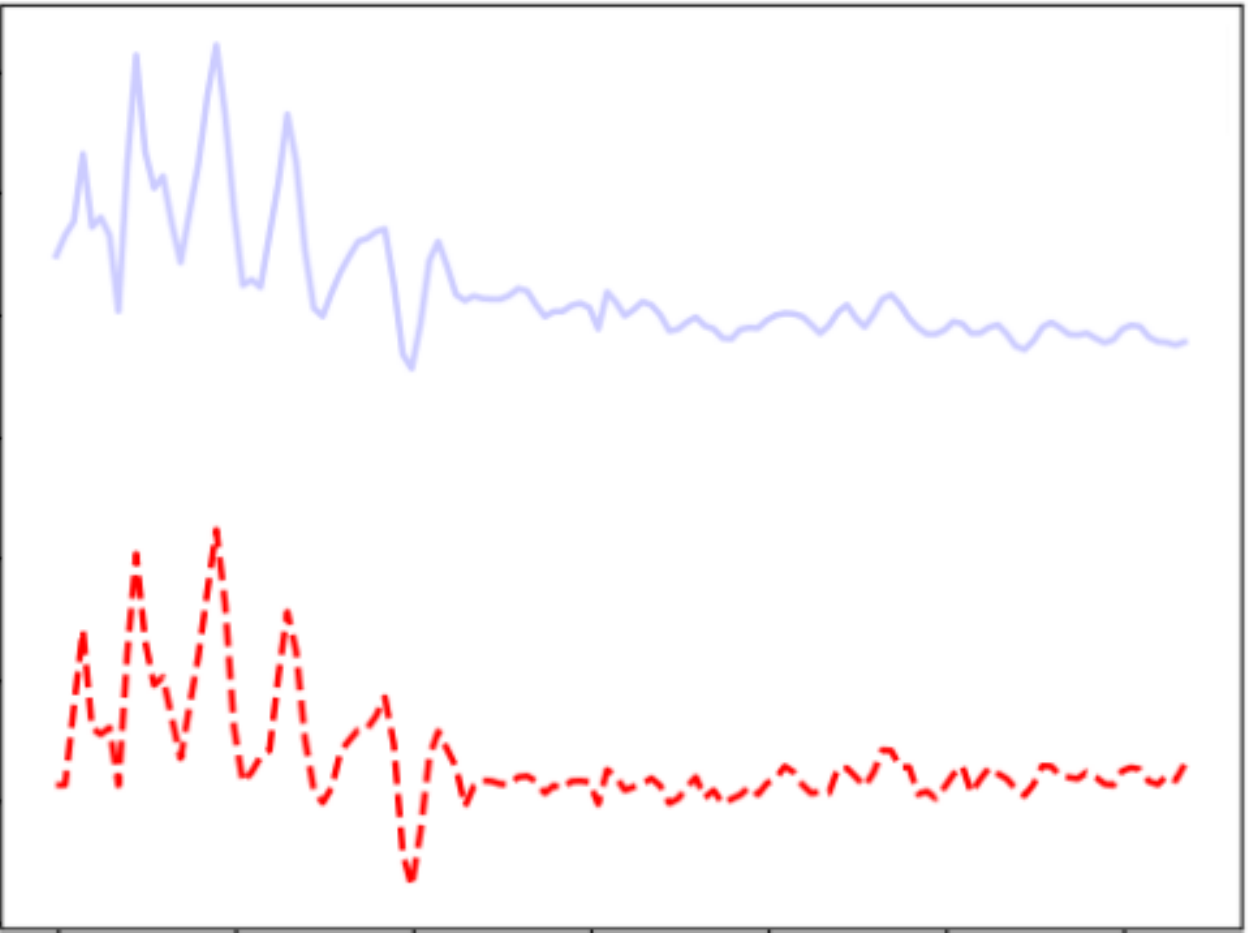}} 
    \subfigure[Situps]{\label{fig:testsample_situps}
        \includegraphics[width=0.48\columnwidth,height=1.15cm]{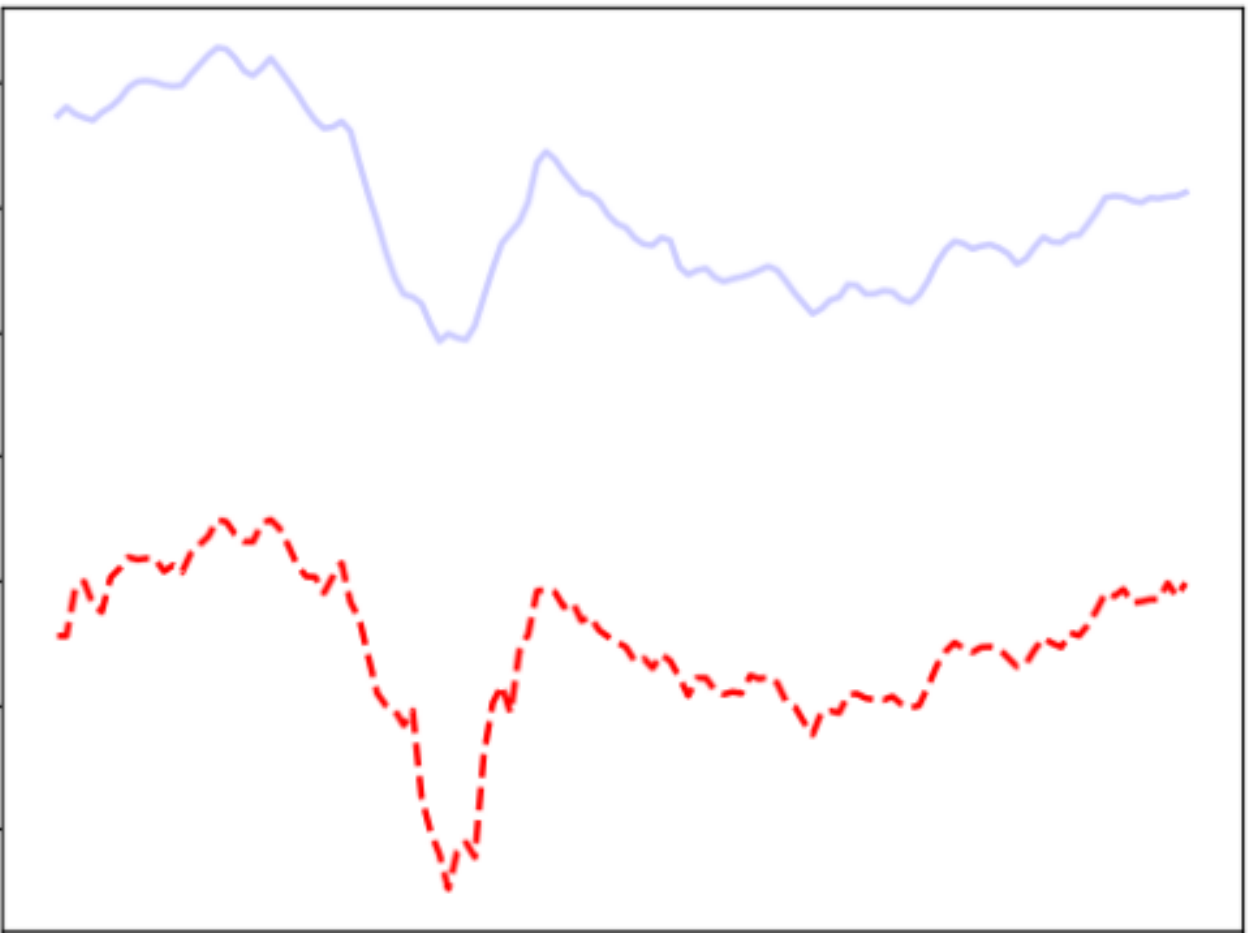}} 

    \subfigure[Dumbbell rows]{\label{fig:testsample_dumbbell_rows}
        \includegraphics[width=0.48\columnwidth, height=1.15cm]{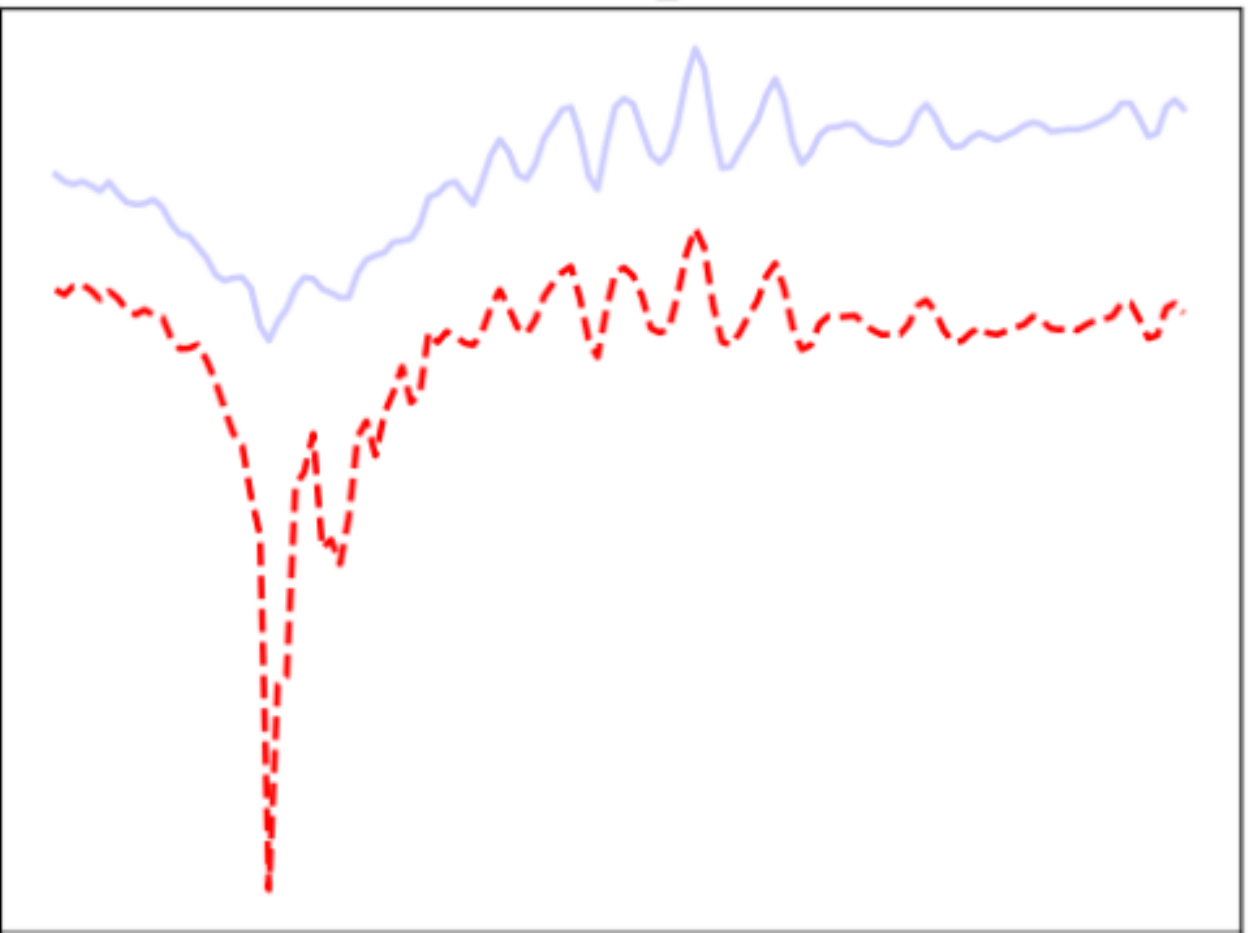}} 
    \subfigure[Dumbbell shoulder press]{\label{fig:testsample_dumbbell_shoulder_press}
        \includegraphics[width=0.48\columnwidth,height=1.15cm]{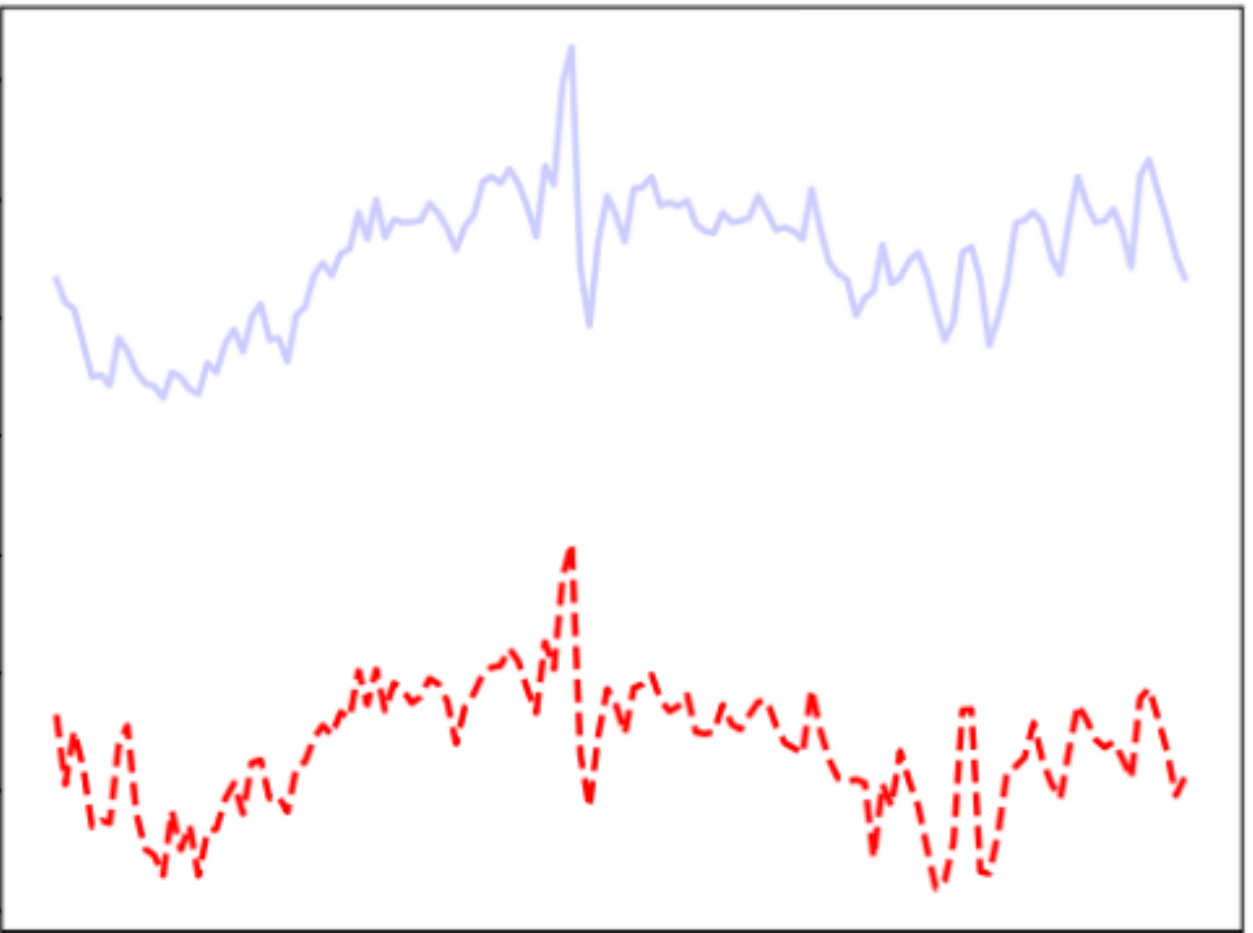}} 

    \subfigure[Tricep extensions]{\label{fig:testsample_tricep_extensions}
        \includegraphics[width=0.48\columnwidth, height=1.2cm]{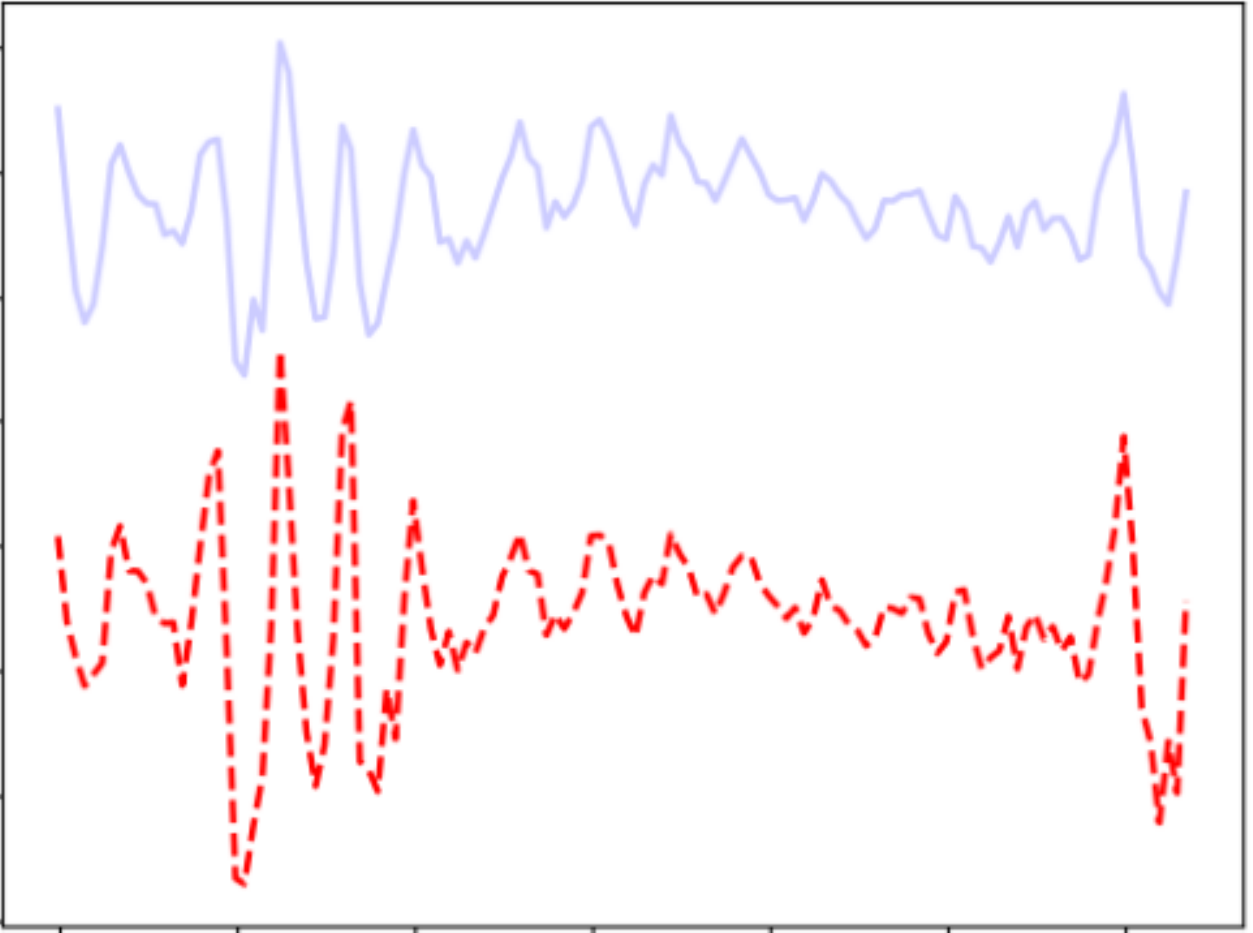}} 
    \subfigure[Lateral shoulder raises]{\label{fig:testsample_lateral_shoulder_raises}
        \includegraphics[width=0.48\columnwidth,height=1.2cm]{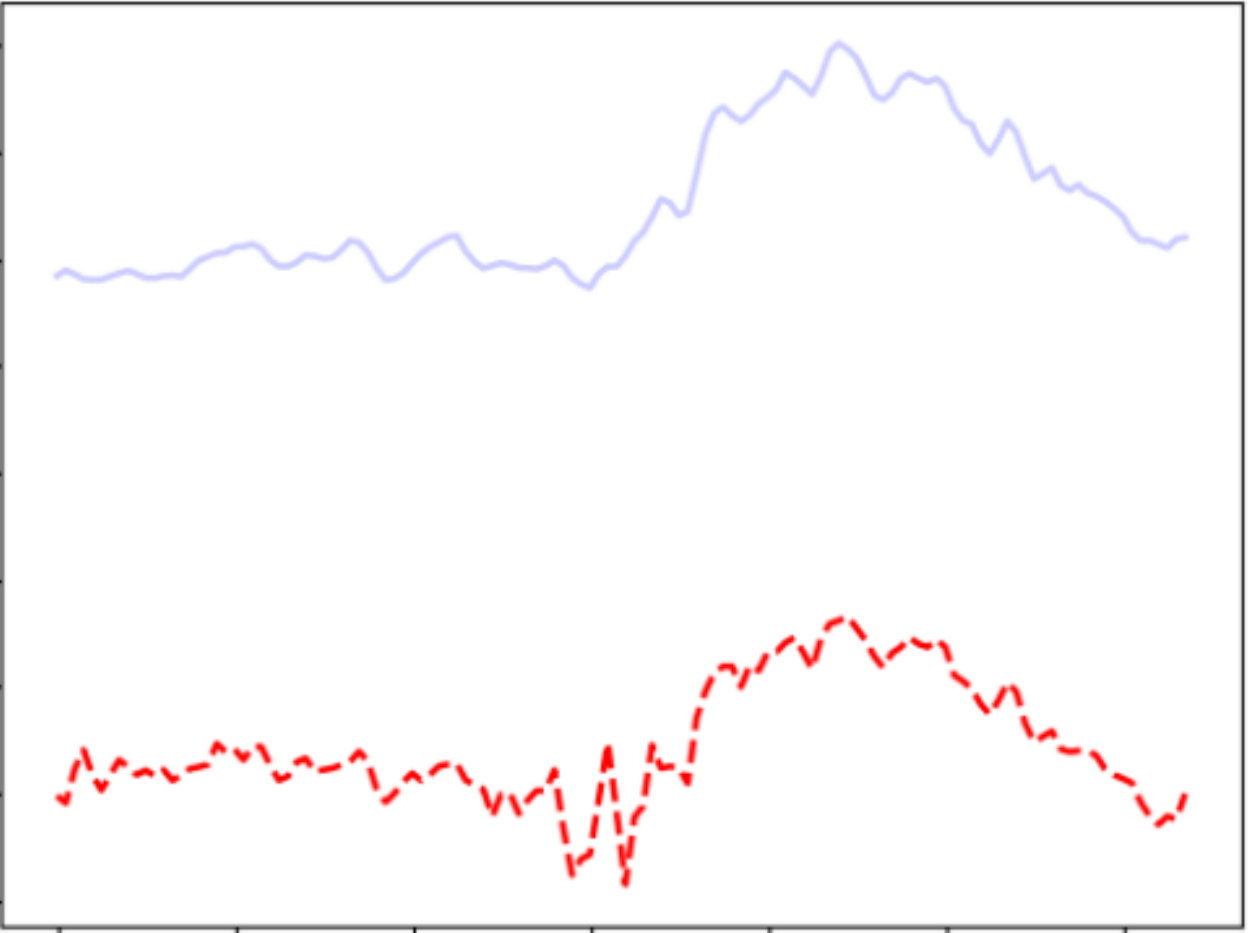}} 
        
    \caption{Examples of generated sensor data (Y-axis: norm acceleration of three axes (\textit{x}, \textit{y}, \textit{z})). The data in red indicates the synthetic sensor data from the diffusion model, while the blue plot represents the real sensor data. 
    }\label{fig:sampleGeneration}
\end{figure} 

\begin{figure}[!t]
    \begin{center}
    \subfigure[\textit{PAMAP2} dataset]{\label{fig:testsample_pamap2}
        \includegraphics[width=0.5\linewidth]{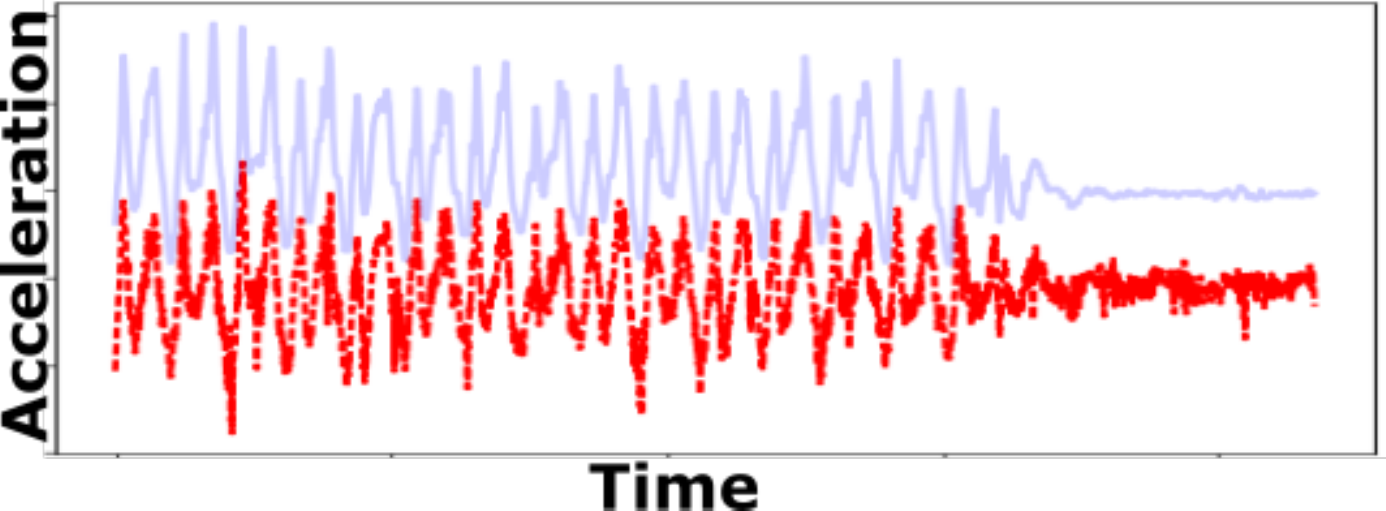}} 
    \hfill
    \end{center}
    \subfigure[lying]{
        \includegraphics[width=0.48\columnwidth, height=1.3cm]{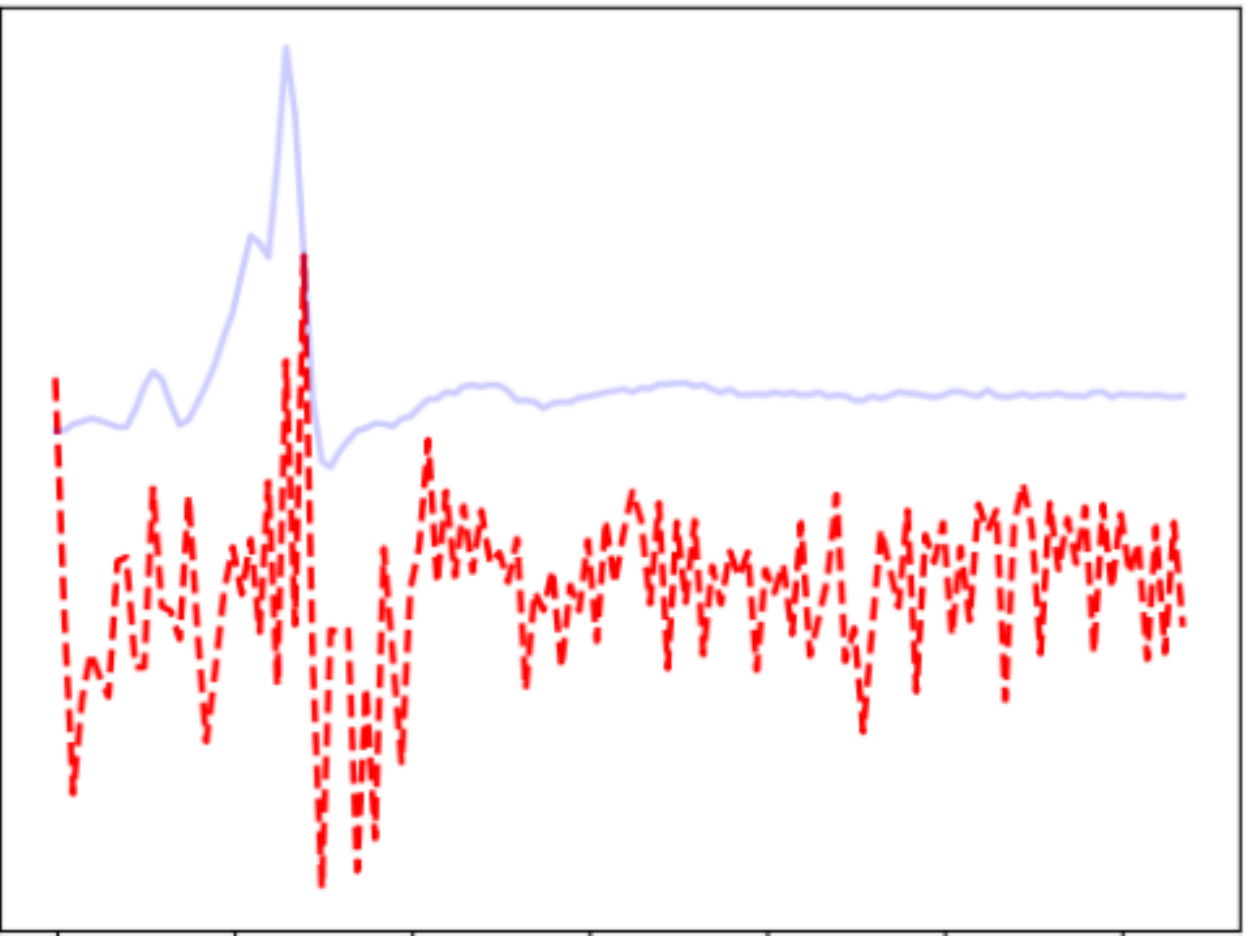}} 
    \subfigure[sitting]{
        \includegraphics[width=0.48\columnwidth,height=1.3cm]{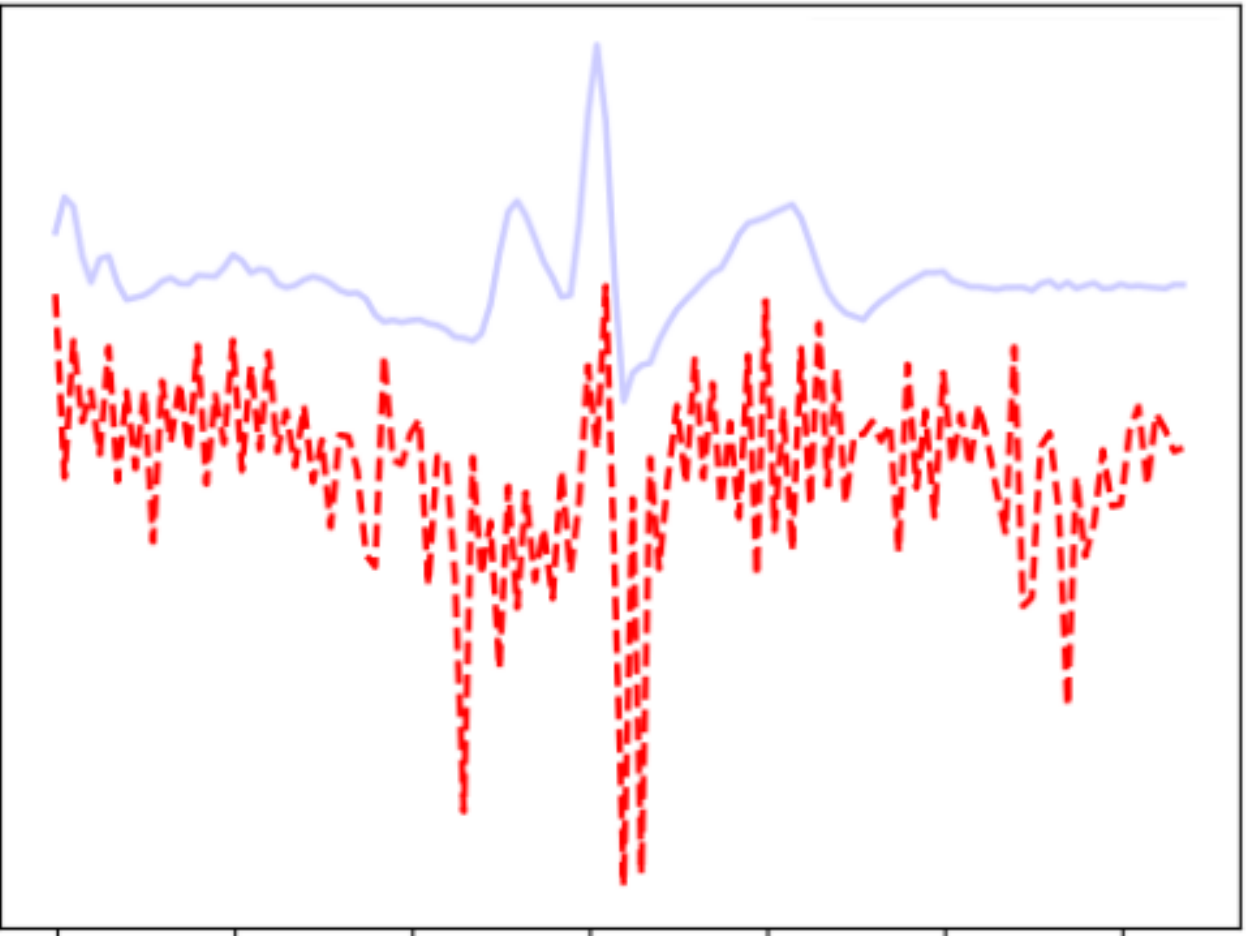}} 

    \subfigure[standing]{
        \includegraphics[width=0.48\columnwidth, height=1.3cm]{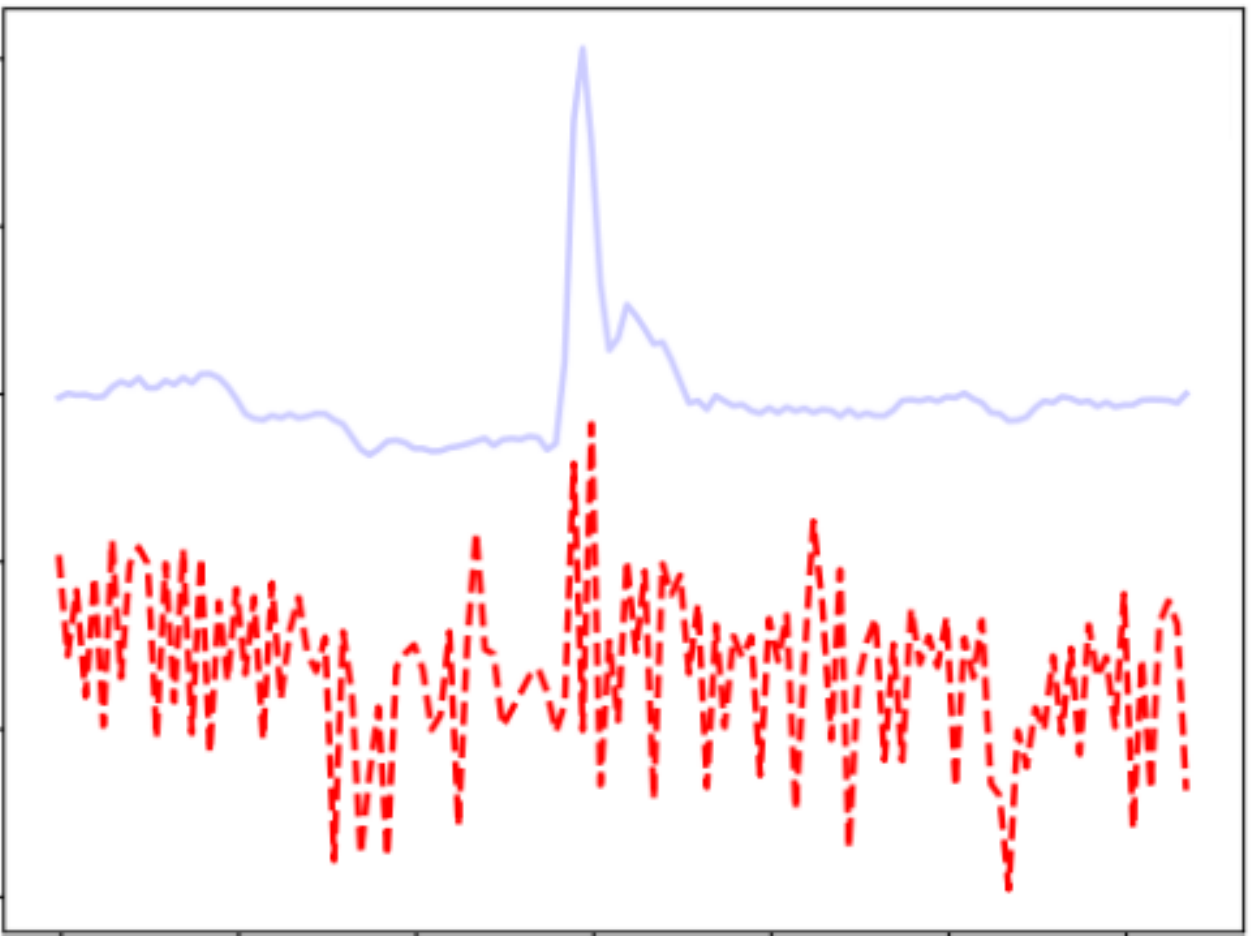}} 
    \subfigure[walking]{
        \includegraphics[width=0.48\columnwidth,height=1.3cm]{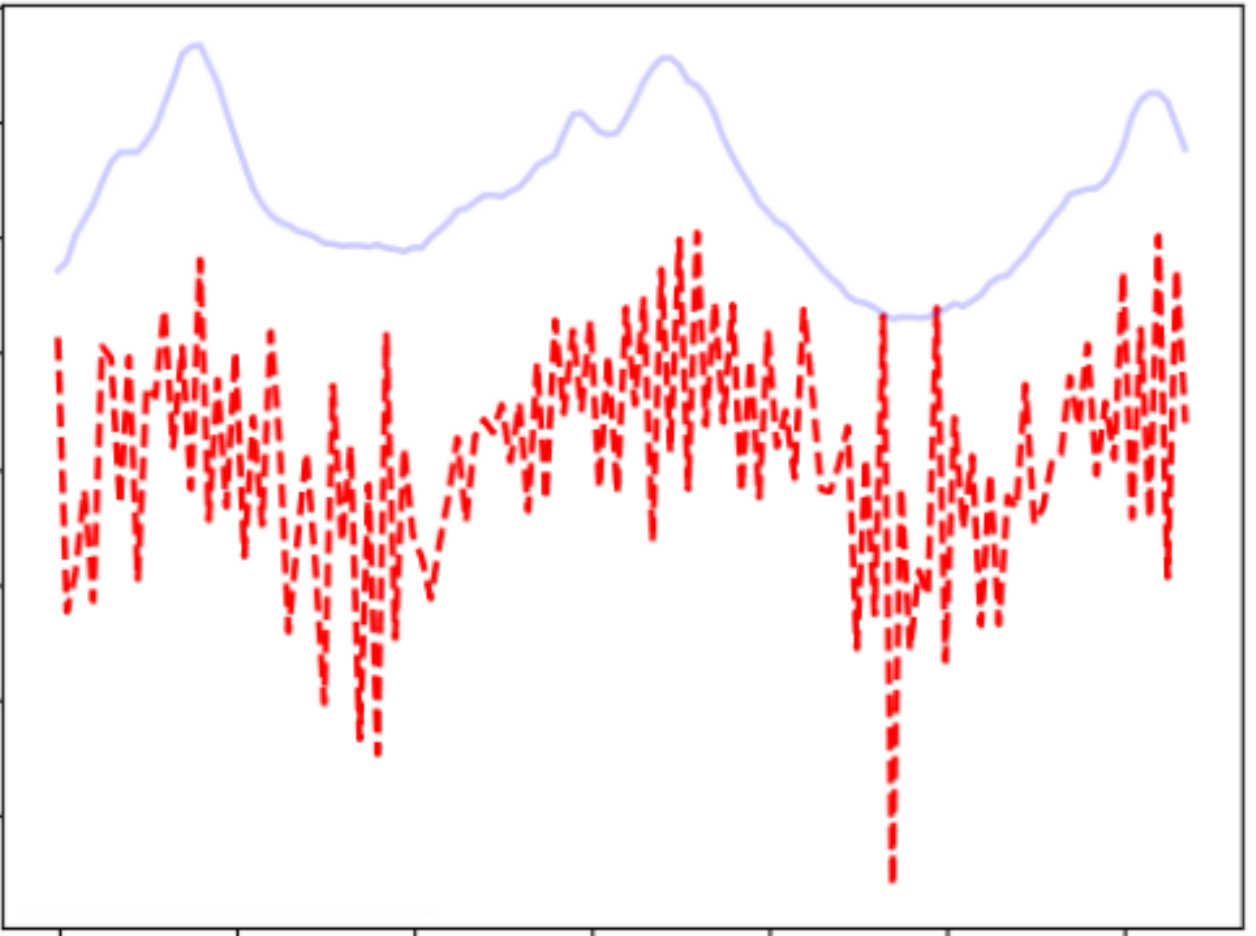}} 

    \subfigure[ascending stairs]{
        \includegraphics[width=0.48\columnwidth, height=1.3cm]{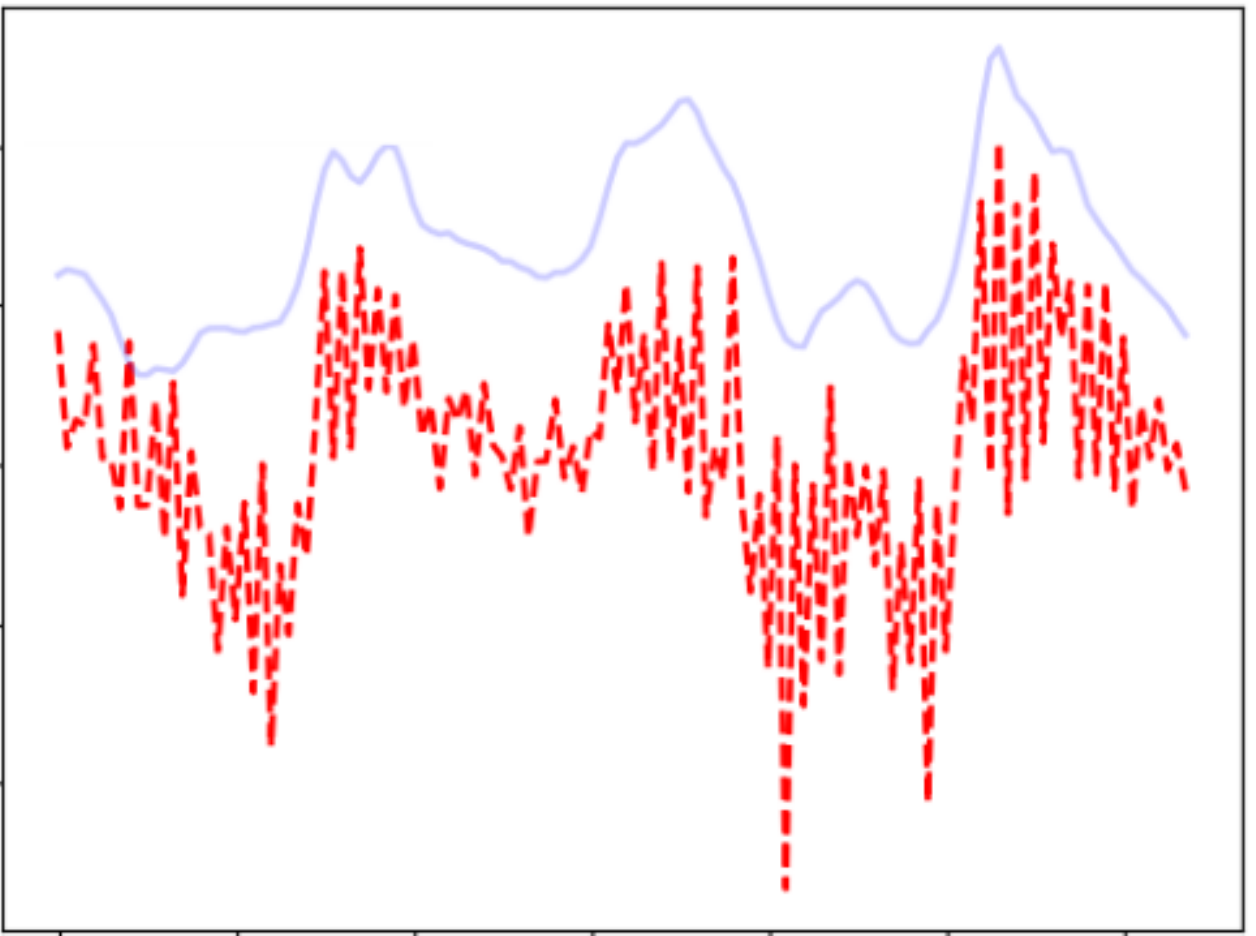}} 
    \subfigure[decending stairs]{
        \includegraphics[width=0.48\columnwidth,height=1.3cm]{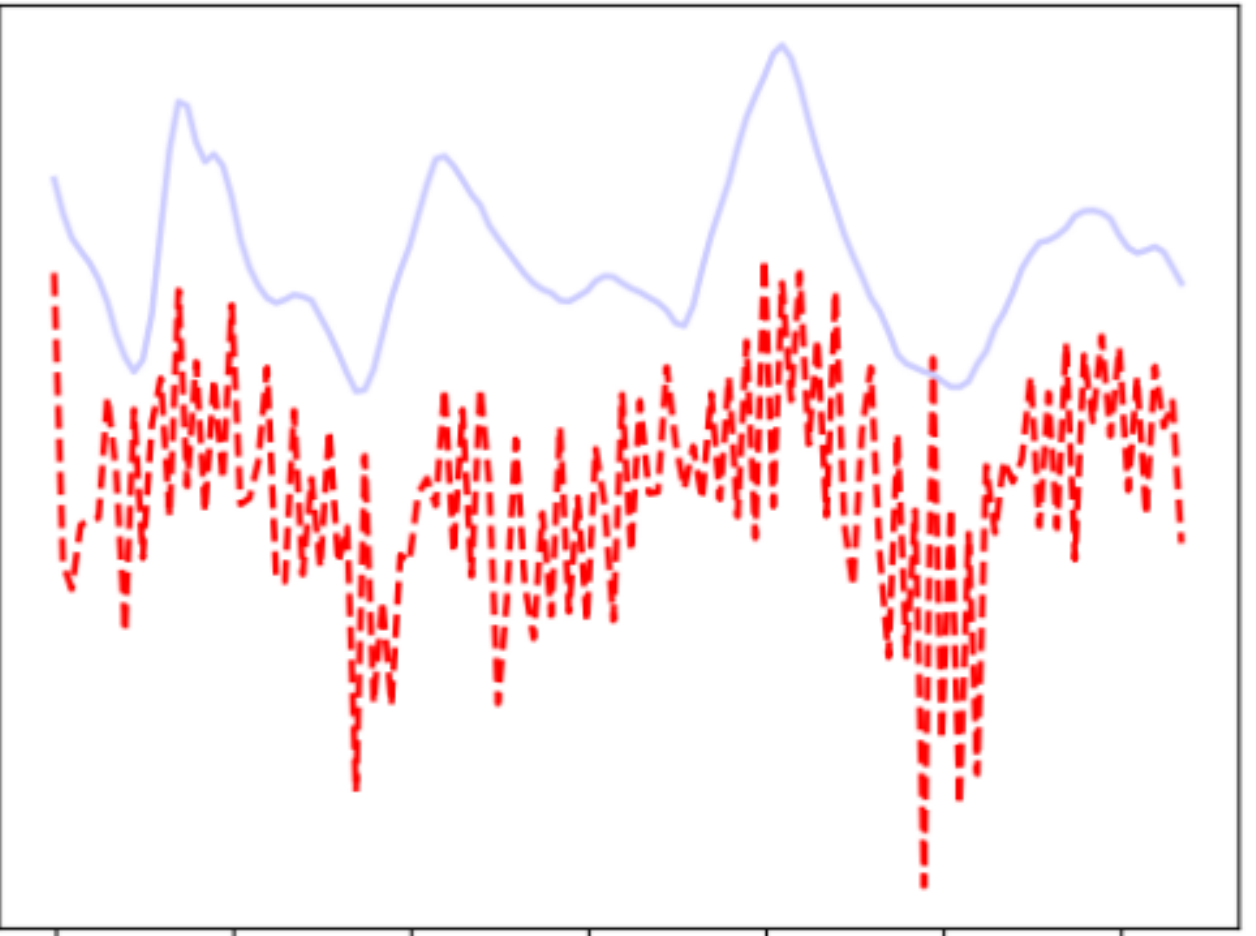}} 

    \subfigure[iroing]{
        \includegraphics[width=0.48\columnwidth, height=1.3cm]{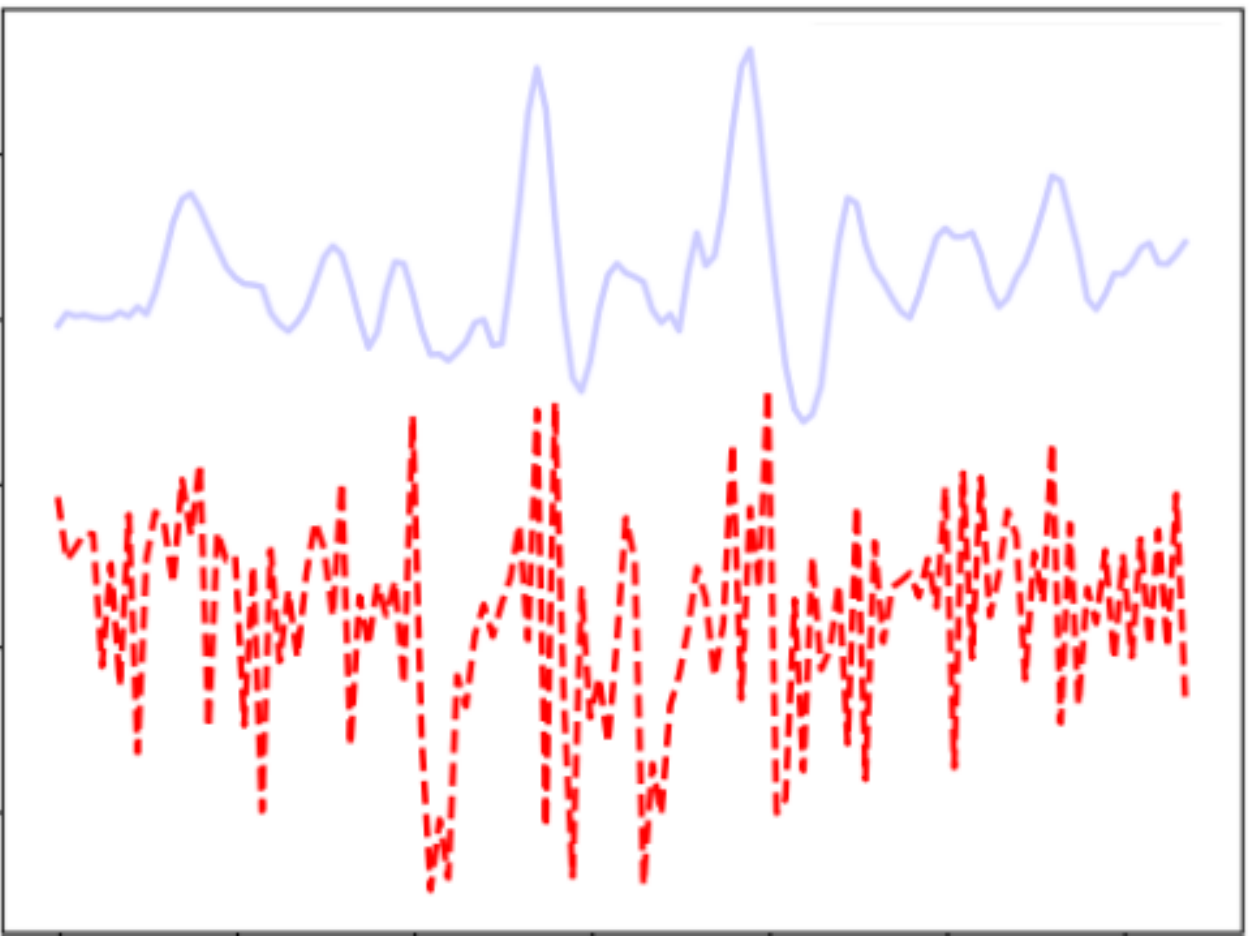}} 
    \subfigure[vacuum cleaning]{
        \includegraphics[width=0.48\columnwidth,height=1.3cm]{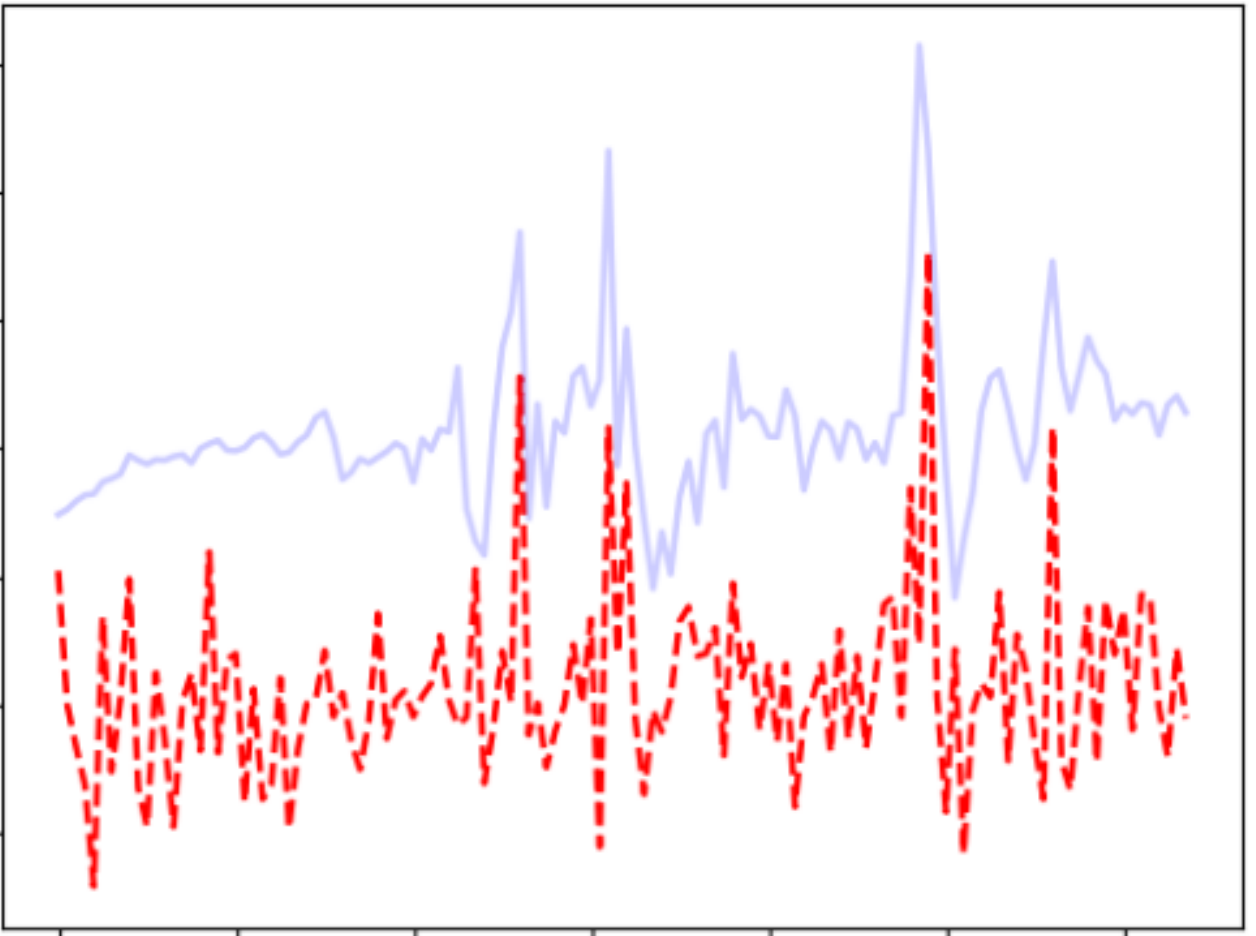}} 
     \vspace{5mm}
    \caption{Examples of generated sensor data (Y-axis: norm acceleration of three axes (\textit{x}, \textit{y}, \textit{z})). The data in red indicates the synthetic sensor data from the diffusion model, while the blue plot represents the real sensor data. 
    }\label{fig:sampleGeneration_pamap2}
\end{figure} 

\begin{figure}[!t]
    \begin{center}
    \subfigure[\textit{Opportunity} dataset]{\label{fig:testsample_oppo}
        \includegraphics[width=0.5\linewidth]{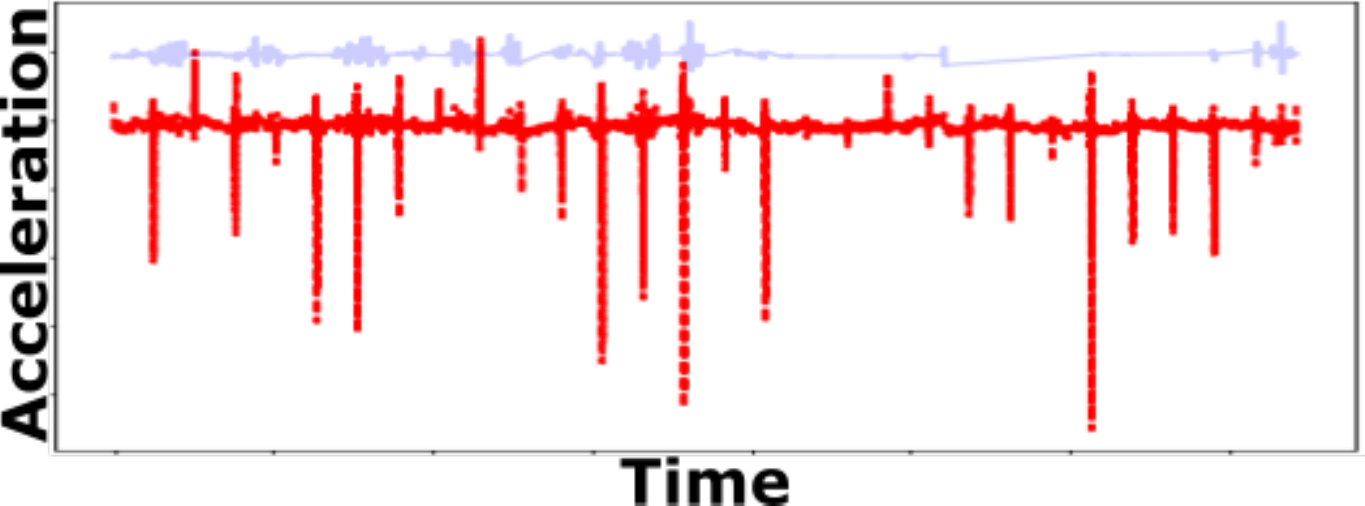}} 
    \hfill
    \end{center}
    \subfigure[Lie]{
        \includegraphics[width=0.48\columnwidth, height=1.3cm]{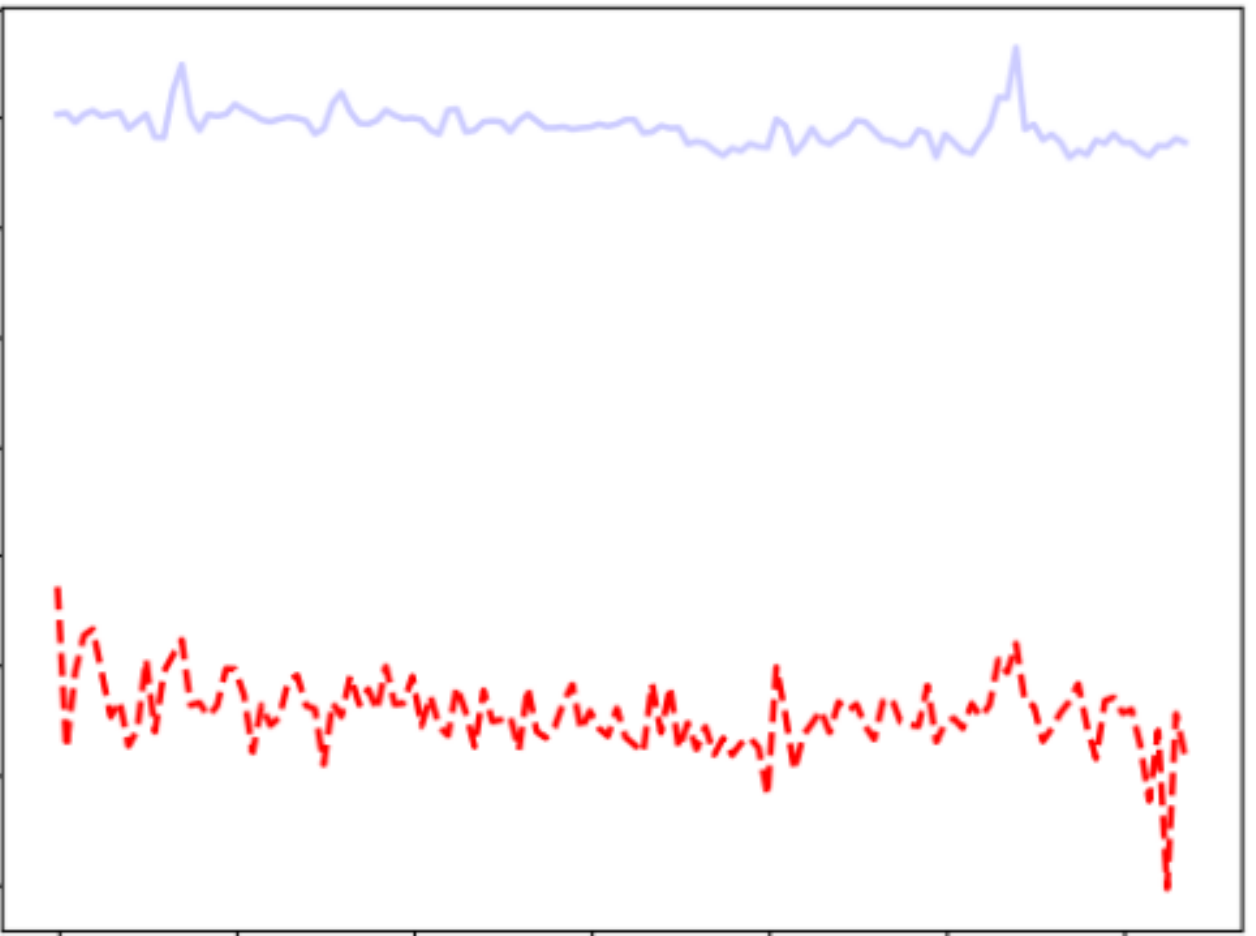}} 
    \subfigure[Sit]{
        \includegraphics[width=0.48\columnwidth,height=1.3cm]{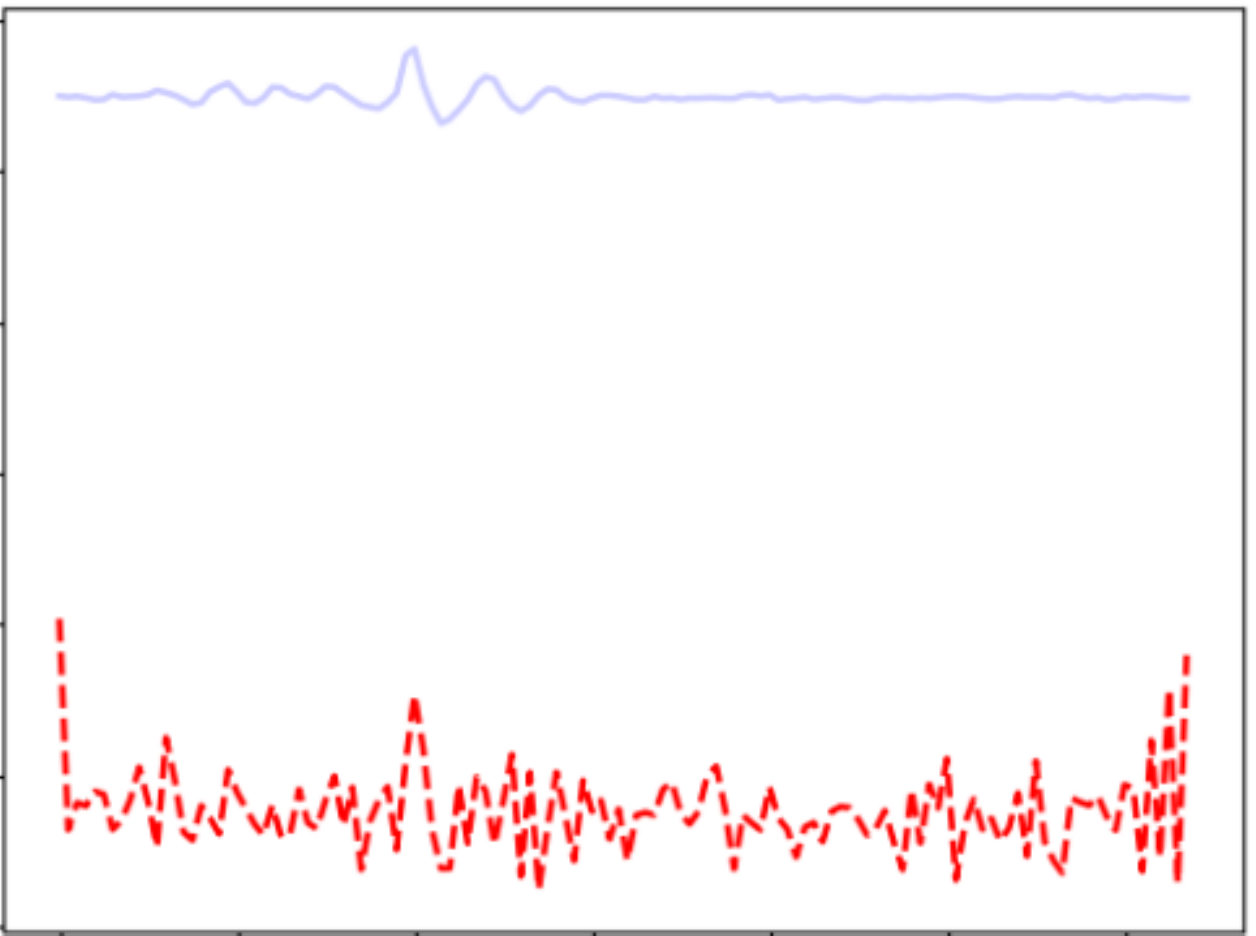}} 

    \subfigure[Stand]{
        \includegraphics[width=0.48\columnwidth, height=1.3cm]{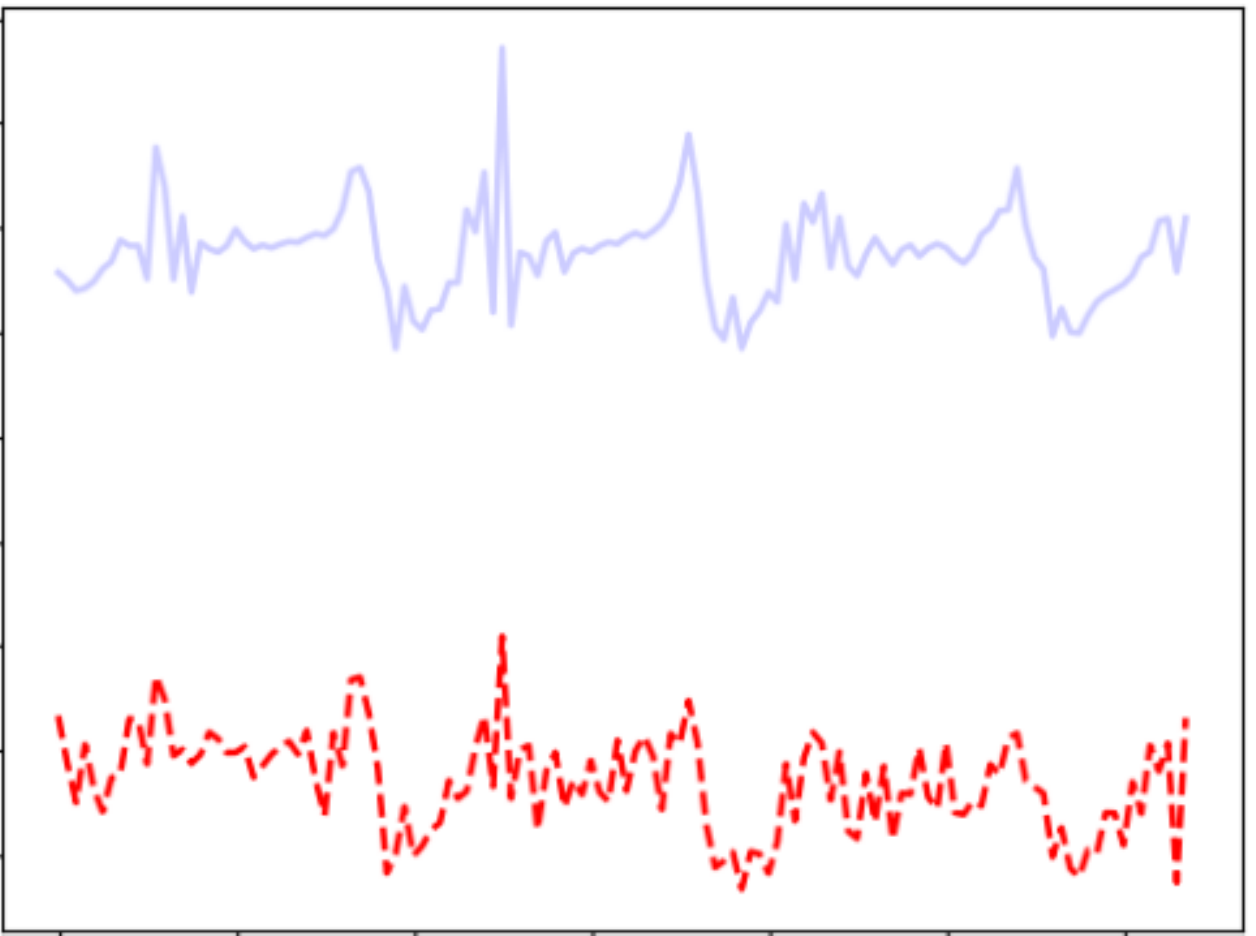}} 
    \subfigure[Walk]{
        \includegraphics[width=0.48\columnwidth,height=1.3cm]{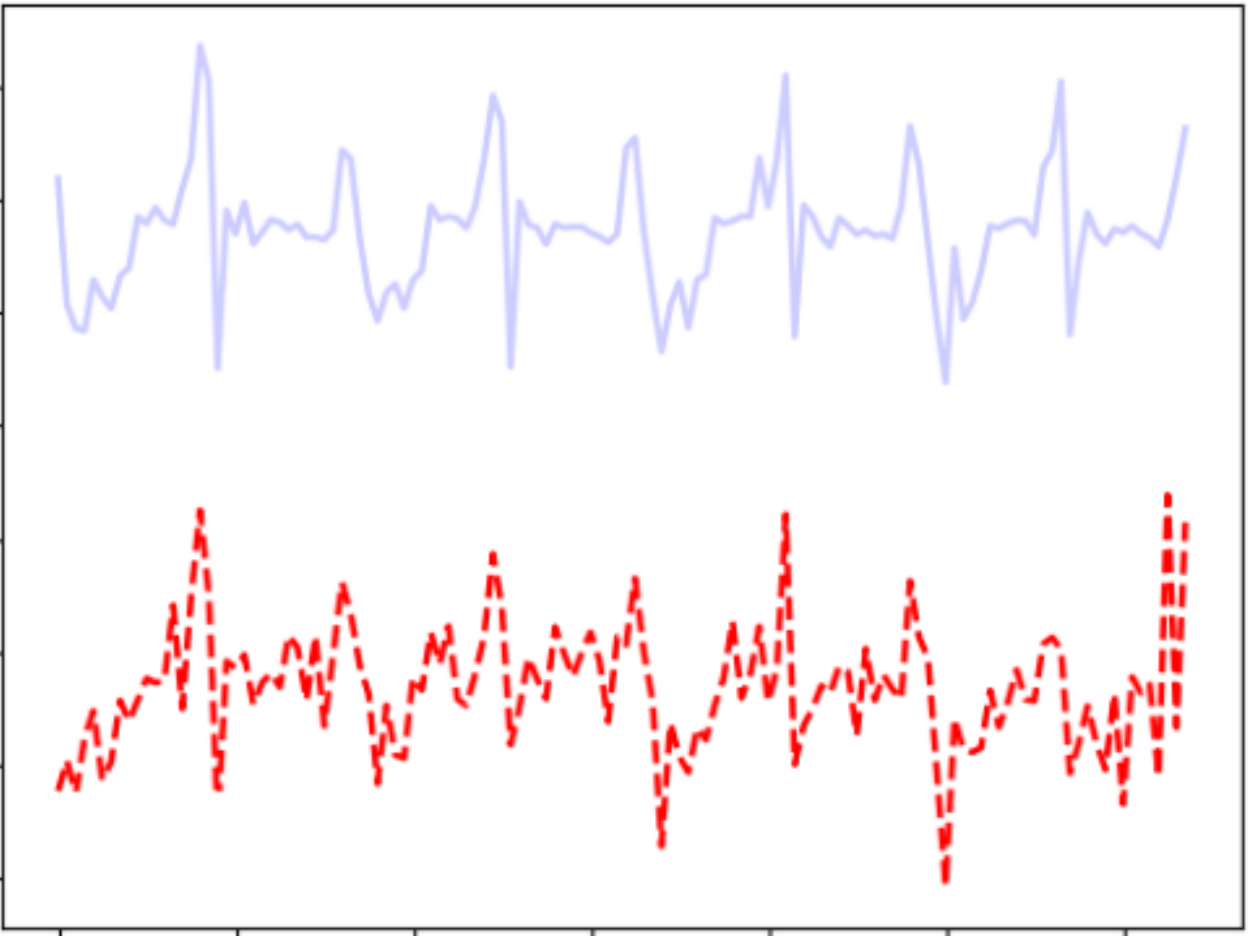}} 

     \vspace{-5mm}
    \caption{Examples of generated sensor data (Y-axis: norm acceleration of three axes (\textit{x}, \textit{y}, \textit{z})). The data in red indicates the synthetic sensor data from the diffusion model, while the blue plot represents the real sensor data. 
    }\label{fig:sampleGeneration_oppo}
\end{figure} 

\subsection{Comparison Results}
Visual examples of generated synthetic sensor data from \textit{MM-Fit}, \textit{PAMAP2} and \textit{Opportunity} datasets are illustrated in \cref{fig:sampleGeneration}, \cref{fig:sampleGeneration_pamap2} and \cref{fig:sampleGeneration_oppo}. \cref{fig:testsample_squats} to \cref{fig:testsample_lateral_shoulder_raises} depict the sample results from different classes in \textit{MM-Fit} dataset. It is worth mentioning that the synthetic sensor data from the model are able to capture the general trends of real data while exhibiting slight variations in finer details across all classes. In the generated results from \textit{PAMAP2} and \textit{Opportunity} datasets (shown in \cref{fig:sampleGeneration_pamap2} and \cref{fig:sampleGeneration_oppo}), although the synthetic data appears to have a higher frequency, it still captures the signal tendencies quite effectively across different classes.

\begin{table}[!t]
  \caption{Comparison of human activity recognition results on different datasets with 20\% 
  }
  \begin{center}
  \renewcommand{\arraystretch}{1.5}
  \resizebox{\textwidth}{!}{%
  \begin{tabular}{cccccccc}
    \hline
      & & \textbf{Baseline} &\textbf{SMOTE} \cite{chawla2002smote}  & \textbf{SVM-SMOTE} \cite{nguyen2011borderline,balaha2023comprehensive}  & \textbf{TimeGAN} \cite{yoon2019time} &\textbf{CC-DM} \cite{shao2023study} &\textbf{SF-DM}
     \\
     \hline
   \multirow{2}{*}{\textit{MM-Fit}} & Accuracy &  0.831$\pm$ 0.0055  & 0.830$\pm$ 0.0065  & 0.836$\pm$ 0.0051 &  0.828$\pm$ 0.0068 & 0.834$\pm$ 0.0072 & \textbf{0.849$\pm$ 0.0072} \\
    & Macro F1   & 0.320$\pm$0.0394 & 0.271$\pm$0.0463 & 0.366$\pm$0.0169 &  0.249$\pm$0.0516 &  0.347$\pm$0.0410 & \textbf{0.386$\pm$0.0495} \\
\hline
       \multirow{2}{*}{\textit{PAMAP2}} & Accuracy &  0.471$\pm$0.0182  & 0.464$\pm$0.0054 & 0.481$\pm$0.0086  &0.482$\pm$0.0081  & 0.464$\pm$0.0085 & \textbf{0.494$\pm$0.0081}  \\
    & Macro F1   & 0.384$\pm$0.0253 & 0.376$\pm$0.0073 & 0.392$\pm$0.0109 &  0.413$\pm$0.0122 & 0.384$\pm$0.0135  & \textbf{0.413$\pm$0.0078} \\
\hline
       \multirow{2}{*}{\textit{Oppo.}} & Accuracy  & 0.408$\pm$0.01951  & 0.495$\pm$0.0165   & 0.482$\pm$0.0244 &  0.449$\pm$0.0301 & 0.430$\pm$0.0234 & \textbf{0.509$\pm$0.0264}\\

        & Macro F1  & 0.258$\pm$0.0303  & 0.364$\pm$0.0191 &  0.340$\pm$0.0369 & 0.329$\pm$0.0436   &  0.276$\pm$0.0433 & \textbf{0.386$\pm$0.0437}\\
   \hline
\end{tabular}}
\end{center}
\label{tab:res20}
\end{table}
\cref{tab:res20} presents comparison results of HAR in terms of accuracy and macro F1 score for several methods on different datasets when only 20\% of the labeled sensor data is available for training. 
Compared to the baseline, the classifier pretrained using our diffusion approach \textbf{SF-DM} consistently yields significant performance improvements in terms of accuracy and macro F1 score across all datasets.
Across all datasets, \textbf{SF-DM} exhibits accuracy improvements ranging from 2.1\% to 24.8\% and macro F1 score improvements from 7.5\% to 49.6\%, compared to the baseline. Specifically, SF-DM achieved an improvement of up to 24.8\% in accuracy score and 49.6\% in macro F1 score compared to the baseline on \textit{Opportunity} dataset. Moreover, in those datasets, we also outperform other oversampling approaches and even TimeGAN by a margin ranging from 1.5\% to 13.4\% in terms of accuracy and with a maximum improvement of 55.0\% in macro F1 score. 

Regarding \textit{Opportunity} dataset, our method significantly outperforms the \textbf{baselines}, \textbf{SMOTE}, \textbf{TimeGAN}, and \textbf{CC-DM} across both accuracy and macro F1 score.

We also highlight the efficiency of training with SF-DM compared to TimeGAN. Since the proposed model is trained in an unsupervised way, there is no need to train separate models for each activity while still maintaining the quality of the generated data. However, with models like TimeGAN, it is essential to have trained models for each class. Take \textit{MM-Fit} dataset as an example, it took around 2 hours 25 minutes for TimeGAN to train a model for a single activity with 20\% of the data while for SF-DM, the training time was 5 hours 45 minutes for all activities on the same platform. Considering the \textit{MM-Fit} dataset includes 10 activities, the total training time for TimeGAN on it would be around 24 hours, which is more than 4 times longer than using the SF-DM.


\subsection{Ablation Study}

\begin{table}[!t]
  \label{tab:acc_macrof1_ablation}
  \setlength\tabcolsep{2pt}
  \begin{center}
  \caption{Ablation study for the diffusion model on \textit{MM-Fit}, \textit{PAMAP2}, and \textit{Opportunity} datasets by changing the proportion of labeled real sensor data.
  }
  \renewcommand{\arraystretch}{1.5}
  \resizebox{\textwidth}{!}{
  \begin{tabular}{cccccccc}
    \hline
    \multirow{2}{*}{proportion} &\multirow{2}{*}{model} & \multicolumn{2}{c}{\textit{MM-Fit}}  &\multicolumn{2}{c}{\textit{PAMAP2}}  &\multicolumn{2}{c}{ \textit{Opportunity}}\\
     \cmidrule(lr){3-4}\cmidrule(lr){5-6}\cmidrule(lr){7-8}
     && \textbf{Acc.} & \textbf{M.F1} & \textbf{Acc.} & \textbf{M.F1} &\textbf{Acc.} & \textbf{M.F1}\\
       
\hline
\multirow{4}{*}{\textbf{0.2}} 
&baseline              & 0.830$\pm$0.0050 &0.322$\pm$0.0498 &0.478$\pm$0.0070 &0.391$\pm$0.0162 &0.421$\pm$0.0276 &0.272$\pm$0.0406 \\
&CC-DM                 & 0.828$\pm$0.0059 &0.321$\pm$0.0593 &0.480$\pm$0.0172 &0.396$\pm$0.0180 &0.391$\pm$0.0229 &0.201$\pm$0.0378\\
& SF-DM[Corresp.P.]    & 0.830$\pm$0.0080 &\textbf{0.353$\pm$0.0641} &0.490$\pm$0.0120 &0.409$\pm$0.0262 &0.467$\pm$0.0313 &0.305$\pm$0.0464 \\
& SF-DM[Proportion: 1] & \textbf{0.834$\pm$0.0026} &0.335$\pm$0.0212 &\textbf{0.492$\pm$0.0141} &\textbf{0.410$\pm$0.0131} &\textbf{0.507$\pm$0.0422} &\textbf{0.416$\pm$0.0771}\\
\hline

\multirow{4}{*}{\textbf{0.3}} 
&baseline             & 0.853$\pm$0.0135  &0.406$\pm$0.0907 &0.499$\pm$0.0085 &0.444$\pm$0.0081 &0.391$\pm$0.0239 &0.215$\pm$0.0442\\
&CC-DM                & 0.867$\pm$0.0142  &0.511$\pm$0.0646 &0.510$\pm$0.0122 &\textbf{0.462$\pm$0.0202} &0.482$\pm$0.0358 &0.362$\pm$0.0514\\
&SF-DM[Corresp.P.]    & 0.852$\pm$0.0040  &0.446$\pm$0.0232 &0.514$\pm$0.0095 &0.452$\pm$0.0081 &0.546$\pm$0.0183 &0.435$\pm$0.0345\\
&SF-DM[Proportion: 1] & \textbf{0.874$\pm$0.0067}  &\textbf{0.517$\pm$0.0210} &\textbf{0.517$\pm$0.0096} &0.453$\pm$0.0141 &\textbf{0.585$\pm$0.0153} &\textbf{0.537$\pm$0.0163}\\
\hline

\multirow{4}{*}{\textbf{0.4}} 
&baseline             & 0.874$\pm$0.0091  &0.530$\pm$0.0559 &0.523$\pm$0.0174 &\textbf{0.467$\pm$0.0239} &0.403$\pm$0.0346 &0.238$\pm$0.0590\\
&CC-DM                & 0.871$\pm$0.0125  &0.524$\pm$0.0489 &0.525$\pm$0.0138 &0.436$\pm$0.0174 &0.496$\pm$0.0392 &0.360$\pm$0.0652\\
&SF-DM[Corresp.P.]    & \textbf{0.882$\pm$0.0071}  &\textbf{0.643$\pm$0.0184} &0.507$\pm$0.0068 &0.453$\pm$0.0113 &0.560$\pm$0.0269 &0.471$\pm$0.0477\\
&SF-DM[Proportion: 1] & 0.879$\pm$0.0014  &0.565$\pm$0.0078 &\textbf{0.525$\pm$0.0128} &0.453$\pm$0.0142 &\textbf{0.568$\pm$0.0446} &\textbf{0.515$\pm$0.0492}\\
\hline
                          
\multirow{4}{*}{\textbf{0.5}} 
&baseline             & 0.857$\pm$0.0186  &0.451$\pm$0.0957 &\textbf{0.520$\pm$0.0057} &\textbf{0.453$\pm$0.0057} &0.456$\pm$0.0459 &0.302$\pm$0.0751\\
&CC-DM                & 0.883$\pm$0.0044  &0.567$\pm$0.0190 &0.495$\pm$0.0135 &0.424$\pm$0.0139 &0.593$\pm$0.0244 &0.511$\pm$0.0386\\
&SF-DM[Corresp.P.]    & 0.877$\pm$0.0114  &0.530$\pm$0.0605 &0.501$\pm$0.0095 &0.445$\pm$0.0146 &0.586$\pm$0.0292 &0.509$\pm$0.0470\\
&SF-DM[Proportion: 1] & \textbf{0.884$\pm$0.0062}  &\textbf{0.567$\pm$0.0121} &0.500$\pm$0.0116 &0.444$\pm$0.0170 &\textbf{0.627$\pm$0.0073} &\textbf{0.573$\pm$0.0164}\\
\hline
                          
\multirow{4}{*}{\textbf{1}} 
&baseline            & \textbf{0.907$\pm$0.0039}  &0.643$\pm$0.0164 &0.551$\pm$0.0094 &0.497$\pm$0.0130 &0.516$\pm$0.0201 &0.408$\pm$0.0346\\
&CC-DM               & 0.901$\pm$0.0040  &0.653$\pm$0.0132 &\textbf{0.556$\pm$0.0137} &0.504$\pm$0.0168 &0.552$\pm$0.0247 &0.453$\pm$0.0409\\
&SF-DM[Corresp.P.]   & 0.882$\pm$0.0062  &0.534$\pm$0.0296 &0.554$\pm$0.0165 &0.471$\pm$0.0217 &0.636$\pm$0.0433 &0.562$\pm$0.0029\\
&SF-DM[Proportion: 1]& 0.906$\pm$0.0085  &\textbf{0.659$\pm$0.0308} &0.554$\pm$0.0140 &\textbf{0.507$\pm$0.0172} &\textbf{0.647$\pm$0.0169} &\textbf{0.598$\pm$0.0115}\\
\hline
\end{tabular}
}
\end{center}
\end{table}
Ablation study results in \cref{tab:acc_macrof1_ablation} indicate the impact of varying proportions of labeled real data for training on the three datasets.  
From the prospect of Macro F1 score, \textbf{SF-DM} [Proportion: 1] outperforms the baseline and other methods in all the cases on the \textit{Opportunity} dataset and most of the cases on other datasets, which highlights the adequacy and effectiveness of statistical features. 

When reducing the amount of unlabeled data available to the diffusion models, we observe decreased improvements and, in some cases, even performance degradation compared to \textbf{SF-DM} [Proportion: 1] or below the baseline. 
This could be attributed to the instability associated with training diffusion models with limited data. When faced with a scarcity of training samples, the model's ability to generalize effectively can be compromised, leading to performance fluctuations and difficulty in accurately capturing the underlying data distribution.  

We also noticed that when the proportion of real data increases from 0.2-0.3 to 0.4-0.5, the performance of the method slightly falls behind the baseline, especially on \textit{PAMAP2} dataset. One possible explanation is that the addition of real data may introduce noise or irrelevant information that adversely affects the model's ability to generalize effectively. As the proportion of real data increases, there is a higher likelihood of encountering outliers or instances that do not match the underlying patterns captured by the model, leading to performance degradation. In this case, we recommend utilizing the method when a small amount of data is available.

Interestingly, in cases where \textbf{SF-DM} shows clearer improvements, such as in \textit{Opportunity}, using only partial data with our statistical information (\textbf{SF-DM} [Corresp: P.]) is overall better than using the label information for diffusion (CC-DM), which can often fall below the baseline results. This indicates the potential of our approach for scenarios with limited labeled data.

Besides, for most of the activities, the model pretrained with synthetic data from SF-DM has a better recognition rate. For data in some categories that have similar signals, for instance, lying, sitting, and standing, synthetic data may not fully capture the subtle differences between these activities, leading to a degradation in performance when the model encounters such instances during testing. 

\begin{figure*}
    \subfigure[\textit{MM-Fit}]{\label{fig:mmfig_ablation }
        \includegraphics[width=0.32\linewidth, height=4.2cm]{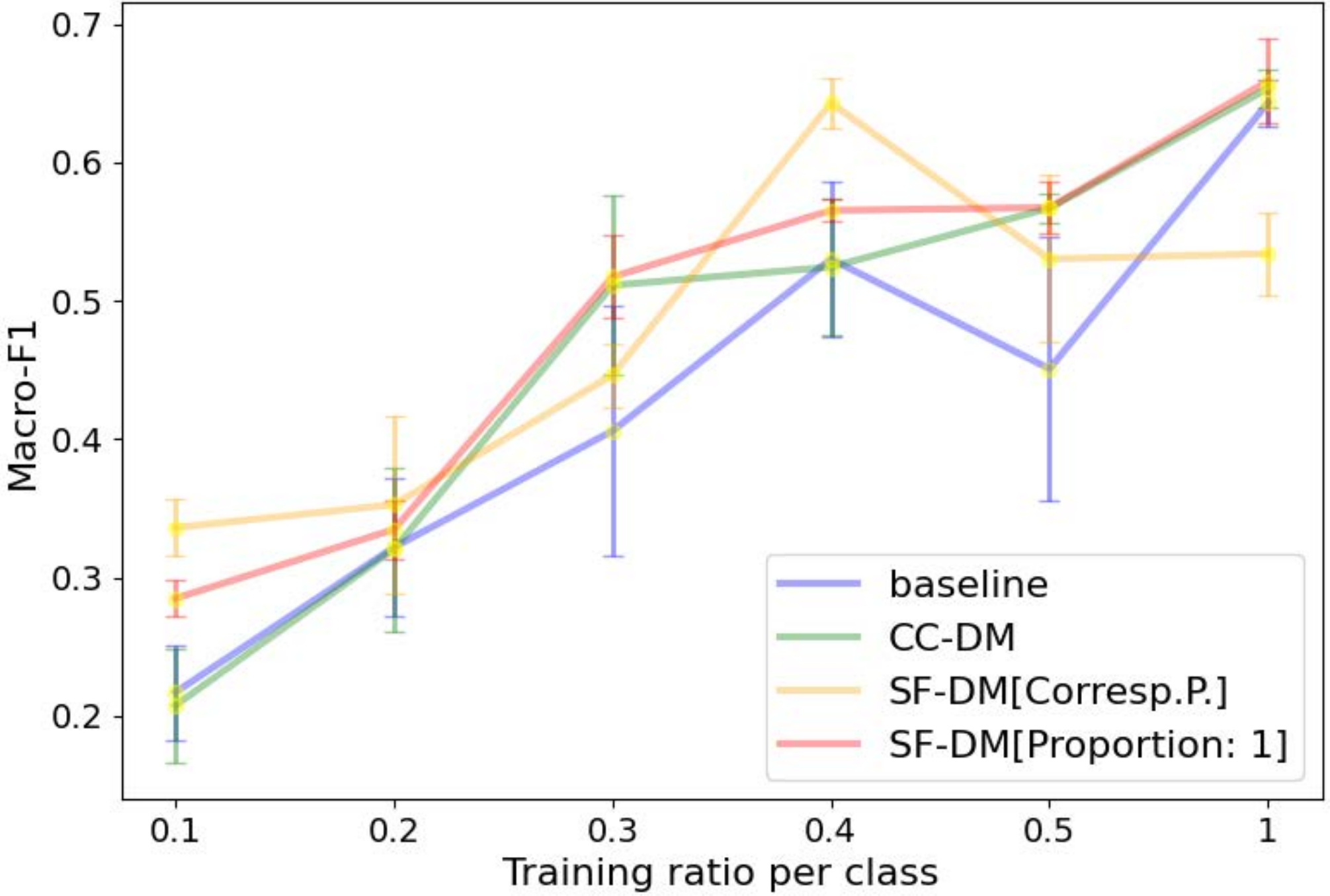}} 
    \subfigure[\textit{PAMAP2}]{\label{fig:pamap_ablation }
        \includegraphics[width=0.32\linewidth, height=4.2cm]{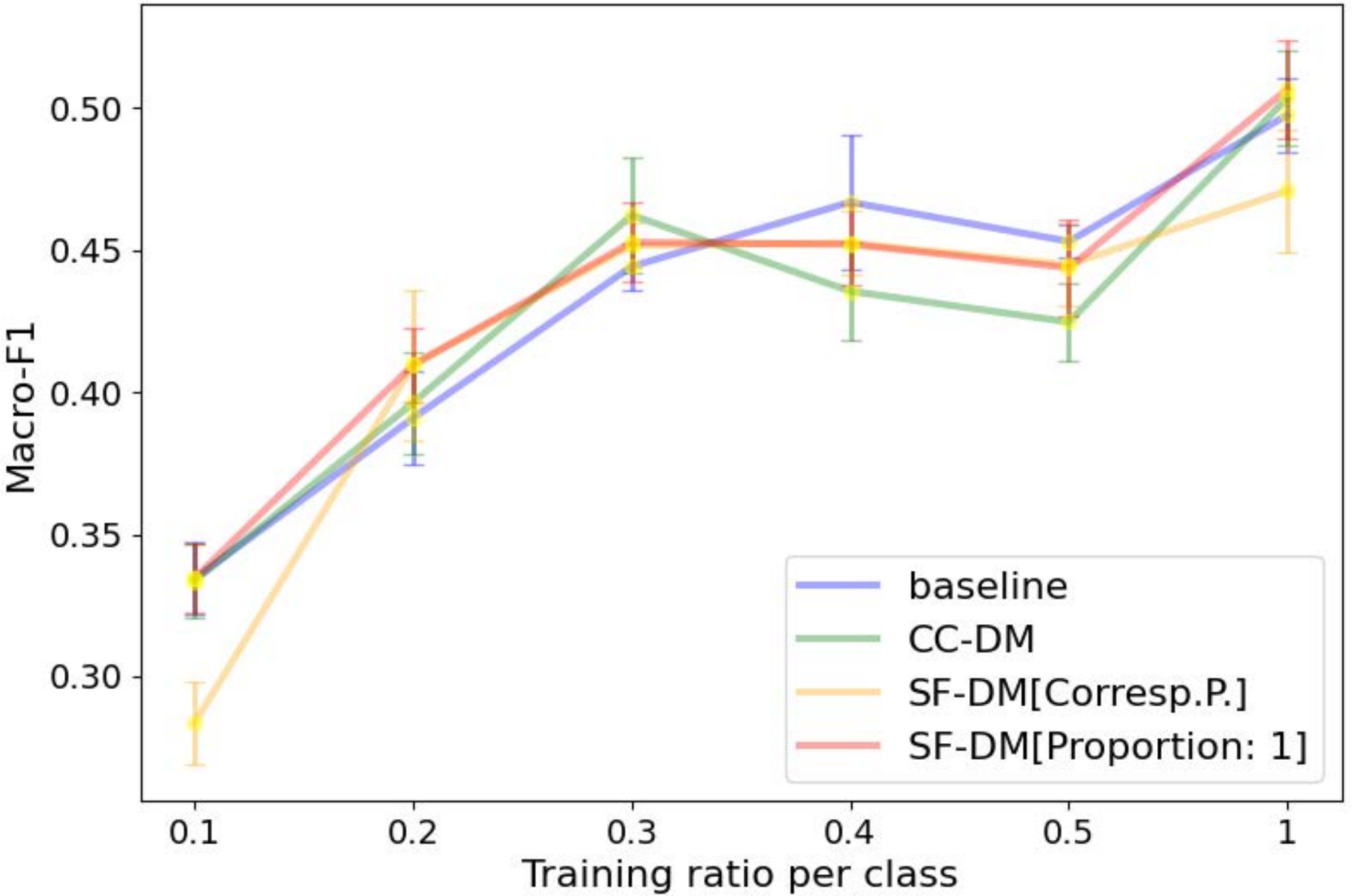}}
    \subfigure[\textit{Opportunity} ]{
        \label{fig:oppo_ablation }
        \includegraphics[width=0.32\linewidth, height=4.2cm]{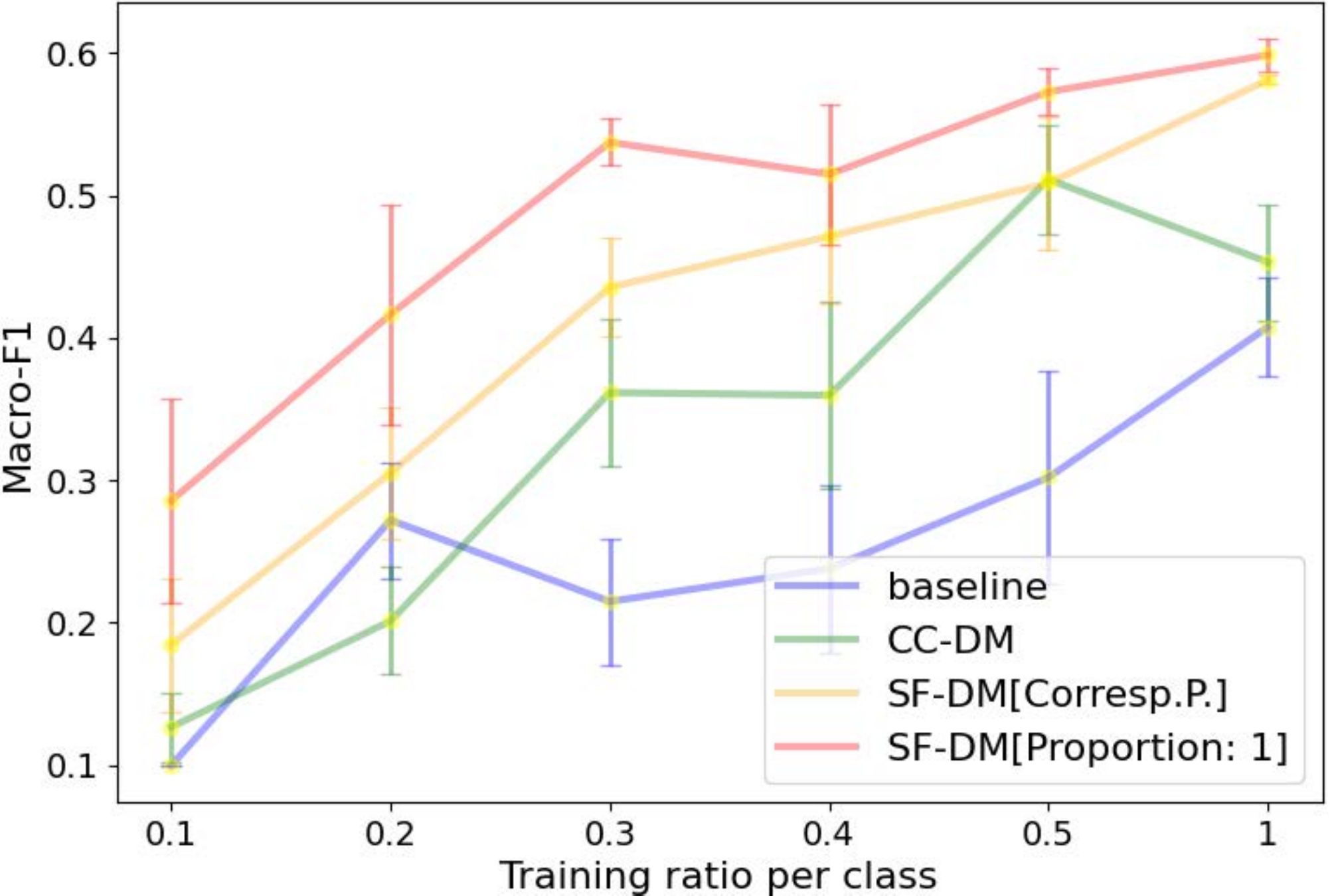}}
        \vspace{-5mm}
        \caption{Ablation study for the proposed diffusion model on \textit{MM-Fit}, \textit{PAMAP2}, and \textit{Opportunity} datasets by changing the proportion of labeled real sensor data. X-axis: proportion of labeled real data usage; Y-axis: Macro F1 score.
}\label{fig:acc_macrof1_ablation}
\end{figure*}

The quantitative evaluation results in terms of Macro F1 score on the three datasets of the ablation study in \cref{tab:acc_macrof1_ablation} are also shown visually in~\cref{fig:acc_macrof1_ablation}. It is clear that increasing the data ratio from 0.1 to 0.3 results in a significant increase in the Macro F1 score for most of the models across all datasets. This indicates that a higher proportion of data contributes positively to the model's performance, potentially allowing it to capture more diverse patterns and improve its ability to generalize across different instances. However, this trend is not consistently observed when continuously increasing the ratio, particularly when transitioning from 0.3 to 0.5. In this interval, the performance of the models varies, indicating that simply increasing the data ratio does not guarantee continued improvement in model performance. This variability suggests the impact of other factors, such as the quality and diversity of the data.

To summarize, the improved performance of the proposed model indicates the capability of SF-DM to capture valuable features from sensor data, guided by selected statistical information. By eliminating the need for labeled data during diffusion model training, we improve the generalizability of the model. As a result, trained SF-DM can be used to generate large amounts of synthetic sensor data that contain the general trends of the real data with slight variations. We also checked the confusion matrices from both the baseline classifier and the classifier pretrained with synthetic data from SF-DM. We noticed that, for most of the categories, the model pretrained with synthetic data from SF-DM has a better recognition rate. For data in some categories that have similar signals, for instance, lie and sit, the classifier pretrained with synthetic data had a lower recognition rate in some trials.

\section{Discussion}
\label{sec:discussion}
To further improve the performance of sensor data generation with the diffusion model and its application in HAR, several approaches can be considered. 

Firstly, the potential for feature diversification is significant. While the accelerometer sensor data has been employed for feature extraction, exploiting information from other modalities could provide richer motion insights. Incorporating RGB and depth data from cameras or other sensor types could offer a more comprehensive view of motion dynamics. This augmentation can capture intricate movement subtleties and improve the quality of synthetic sensor data. Furthermore, the scope of the diffusion model is not confined to single-modality data. Extending the model's capability to generate multi-modal data is a promising step towards robust HAR enhancement. Integrating data from diverse sources like audio, video, or even textual descriptions can significantly have the potential to enhance the quality and realism of synthetic sensor data, improving the understanding of complex human activities, and facilitating more accurate and robust analysis in applications such as human activity recognition and motion tracking.

Secondly, the current model \textbf{SF-DM} can be extended to generate latent representations rather than raw sensor data. This strategy leverages the diffusion model's ability to learn shared and relevant features across multiple modalities. By training different autoencoders, the latent representation obtained from the diffusion model can be effectively translated into various data modalities by simply providing modality-specific information. This extension enriches the diversity and effectiveness of learned representations, potentially leading to more powerful models for HAR.

Furthermore, unlike vision-based data, there are fewer useful metrics available for evaluating the quality of sensor data. Due to the nature of sensor data, factors such as temporal dynamics, signal noise, and sensor fusion should be considered. Novel evaluation metrics can be devised to capture the nuanced characteristics of HAR datasets (e.g., data from lying, sitting, and standing). The development and utilization of appropriate evaluation metrics specifically tailored to assess the quality and validity of sensor data generated in the context of HAR enable better assurance of measurement accuracy and validation of performance improvements.


\section{Conclusion}
\label{sec:conclusion}
In this paper, we proposed a novel unsupervised statistical feature-guided diffusion model for sensor-based HAR. Operating within an encoder-decoder framework, our diffusion model employs statistical features, including mean, standard deviation, Z-score, and skewness, to generate high-quality sensor data without relying on class label information. 
Quantitative evaluations demonstrate the model's consistent superiority over traditional oversampling methods and TimeGAN, with accuracy improvements ranging from 2.1\% to 24.8\% and macro F1 enhancements of 7.5\% to 49.6\%. Visually, the synthetic sensor data generated by our model were visually scrutinized and found to faithfully capture the essence of real data, albeit with subtle variations in detail across different classes.

The significance of our work extends beyond HAR, as our approach has the potential to be applied in various domains that require time-series data, especially in scenarios with limited labeled data. 
In future work, we aim to optimize the training procedure of our framework and utilize class distribution information. 
In addition, we plan to extend it to sensor data generation from other modalities, such as video data, and explore additional statistical features from both the time and frequency domains.


\bibliographystyle{ACM-Reference-Format}
\bibliography{ref}


\end{document}